\documentclass[%
 reprint,
superscriptaddress,
 amsmath,amssymb,
 aps, prb,
floatfix,
]{revtex4-1}
\usepackage[normalem]{ulem}
\usepackage{graphicx}
\usepackage{dcolumn}
\usepackage{bm}
\usepackage{color}
\begin{document}


\title{The strain-stress relationships for coherent in-plane strain in heterostructures with monoclinic crystal systems: $\beta$-(Al$_x$Ga$_{1-x}$)$_2$O$_3$ on $(h0l)$ $\beta$-Ga$_2$O$_3$ as example}

\author{Mathias Schubert}
\email{schubert@engr.unl.edu}
\homepage{http://ellipsometry.unl.edu}
\affiliation{Department of Electrical and Computer Engineering, University of Nebraska-Lincoln, Lincoln, NE 68588, USA}
\affiliation{NanoLund and Solid State Physics, Lund University, 22100 Lund, Sweden}
\author{Rafa\l{} Korlacki}
\affiliation{Department of Electrical and Computer Engineering, University of Nebraska-Lincoln, Lincoln, NE 68588, USA}
\author{Vanya Darakchieva}
\affiliation{Terahertz Materials Analysis Center and Center for III-N technology, C3NiT -- Janz\`{e}n, Department of Physics, Chemistry and Biology (IFM), Link\"{o}ping University, 58183 Link\"{o}ping, Sweden}
\affiliation{NanoLund and Solid State Physics, Lund University, 22100 Lund, Sweden}

\date{\today}

\begin{abstract}
In this work we derive the state of strain or stress under symmetry conserving conditions in pseudomorphic lattices with monoclinic symmetry. We compare surface vectors across the template epitaxial layer interface and impose conditions of a stress free epitaxial layer. As a result, we demonstrate the existence, in theory, of exactly three possible unit cells which can establish onto a given template. We demonstrate this approach for a class of templates with $(h0l)$ planes and $\beta$-(Al$_x$Ga$_{1-x}$)$_2$O$_3$ on $(h0l)$ $\beta$-Ga$_2$O$_3$. We discuss the effects of composition $x$ and surface orientation onto the formation of three elastically stable unit cells, their strain and stress tensors, unit cell axes, unit cell volumes, lattice spacing, elastic potential energies, and stress free directions. The previous paradigm for epitaxial layer growth where the stress free direction is always perpendicular to the growing surface is not generally valid for low symmetry materials. In the example here, we find two possible competing domains with stress free direction oblique to the surface of the template for almost all planes $(h0l)$. We calculate the band-to-band transitions for $\beta$-(Al$_{0.1}$Ga$_{0.9}$)$_2$O$_3$ on $(h0l)$ $\beta$-Ga$_2$O$_3$ using the composition dependent deformation parameters and elastic coefficients reported prevoiously [Korlacki~\textit{et al.} Phys. Rev. Appl.~\textbf{18}, 064019 (2022)].
\end{abstract}

\maketitle

\section{Introduction}
In heteroepitaxial growth of pseudomorphic single crystals onto single crystal templates the mismatch between lattice constants of isomorphic unit cells leads to biaxial strain along the surface of the template. The strain results in a characteristic stress distribution across the epitaxial layer. The effects of strain and stress are crucially important for understanding of modifications in thermal, electronic, optical, and optoelectronic properties, for example. The full evaluation of the strain-stress relationships and the consequences for stable growth of single domain epitaxial layers has not been elucidated for low symmetry crystals yet. Low-symmetry materials such as monoclinic $\beta$-Ga$_2$O$_3$ and the heteroepitaxy with isostructure and isovalent alloys $\beta$-Ga$_2$O$_3$ and $\beta$-Ga$_2$O$_3$ is of high contemporary interest due to various emergent physical properties such as a very large tunable band gap energy. For example, isovalent alloying of In and Al in $\beta-$(Al,Ga,In)$_2$O$_3$ offers tuning of the band to band transition energies using composition and lattice-mismatch induced strain.\cite{doi:10.1021/acsami.3c19095} Monoclinic $\theta$-Al$_2$O$_3$ has a calculated lowest band to band transition of 7.2–7.5~eV\cite{ PhysRevApplied.18.064019, 10.1063/1.5036991}and monoclinic $\beta$-In$_2$O$_3$ of 2.7 eV.\cite{10.1063/1.5093195, PhysRevB.92.085206} with lattice mismatch to $\beta$-Ga$_2$O$_3$ of 4\%\cite{Ahman:fg1144, 10.1063/1.4913447} and 10\%.\cite{10.1063/5.0078037} In heteroepitaxial growth strain and stress are inherent due to thermal expansion and lattice mismatch between heterostructure constituents. Modifications imposed by strain onto band structure properties of crystalline materials can be conveniently studied using group theoretic methods.\cite{ BirPikusBook} When lattice distortions are not too large, a range of distortions can exist where the resulting variations of band structure properties can be approximated as a linear perturbation of the eigenstates of the strain-free lattice. Then, deformation parameters can be used to calculate the effect of a given eigenstate as a result of a given condition of the lattice's strain or stress. The set of equations rendering this relationship are also known as strain-stress relationships for eigenstates in crystals. These structure of the set of equations is conditioned by the symmetry of the participating lattices within a given heterostructure.

Zhang~\textit{et al.} performed density functional theory analysis of hydrostatic strain induced variations in the band structure of $\beta$-Ga$_2$O$_3$ and reported strong variation in band to band transitions, anisotropy of electron mobility, and effective mass among others. The strong variations suggest strain engineering as an important approach to band gap engineering in $\beta$-Ga$_2$O$_3$.\cite{ZHANG2023414851}
Huang~\textit{et al.} reported on the $\beta$ to $\gamma$ phase transformation in Sn-doped and Si-implanted Ga$_2$O$_3$ as a result of local strain induced by impurity atoms which favor formation of interstitial-divacancy complexes with subsequent phase transition.\cite{ 10.1063/5.0156009}  Seacat, Lyons and, Peelaers predict $\beta$-(In$_{0.25}$Al$_{0.75}$)$_2$O$_3$ as composition with lattice constants similar to $\beta$-Ga$_2$O$_3$ with indirect/direct band gap of 5.96~eV/5.7~eV and a conduction band offset of 1~eV towards $\beta$-Ga$_2$O$_3$.\cite{PhysRevMaterials.8.014601} Barmore~\textit{et al.} investigated the effects of hydrostatic pressure up to 9~GPa on the photoluminescence shift of the R lines in Cr-doped $\beta$-Ga$_2$O$_3$  and (Al$_{0.1}$Ga$_{0.9}$)$_2$O$_3$.\cite{10.1063/5.0149900} Xu~\textit{et al.} demonstrated strain-induced phase transformation and stabilization under large template mismatch between rutile structure $\alpha$-Al$_2$O$_3$  and $\beta$-Ga$_2$O$_3$ \cite{ doi:10.1021/acsami.8b17731} Zhang~\textit{et al.} used density functional theory computations and discussed the effect of uniaxial, biaxial, and isotropic strain onto the band gap energy and effective mass parameter in $\beta$-Ga$_2$O$_3$.\cite{ ZHANG2023106916} Hara~\textit{et al.} performed Vicker’s indentation with Raman studies on (010) $\beta$-Ga$_2$O$_3$ and deduced deformation potential parameters for selected modes.\cite{Hara_2023} Uchida and Sugie reported stress analysis in $\beta$-Ga$_2$O$_3$ using Raman spectroscopy and obtained phonon deformation potential parameters in very good agreement with theoretical predictions by Korlacki~\textit{et al.}\cite{Uchida_2023, PhysRevB.102.180101} Hasuike~\textit{et al.} performed bending stress investigations using Raman spectroscopy bending (001) $\beta$-Ga$_2$O$_3$ along the [001] direction thereby also deforming the monoclinic lattice cell.\cite{ Hasuike_2023} The resulting deformation potential parameters assuming monoclinic symmetry differ from those predicted for symmetry-retaining stress.\cite{ PhysRevB.102.180101} Mu~\textit{et al.} calculated the orientation-dependent band offsets in pseudomorphically strained and relaxed $\beta$-(Al,Ga)$_2$O$_3$/$\beta$-Ga$_2$O$_3$ interfaces.\cite{10.1063/5.0036072} respectively.\cite{10.1063/1.5093195, Ahman:fg1144} Korlacki~\textit{et al.} described the linear perturbation theory strain and stress relationships for optical phonon modes in monoclinic crystals for strain and stress situations which maintain the monoclinic symmetry of the crystal.\cite{ PhysRevB.102.180101} The relationships were extended for the $\Gamma$-point topmost valence band and lowest conduction band levels for $\beta$-Ga$_2$O$_3$ and monoclinic $\theta$-Al$_2$O$_3$.\cite{PhysRevApplied.18.064019}

Key to correct interpretation of the state of strain in a given heteroepitaxial or otherwise strained system is the establishment of correct relationships between the implied stress and the resulting strain. This task is not trivial for low-symmetry crystals. Grundmann provided analytic solutions for the calculation of the strain parameters in pseudomorphically grown epitaxial layers of monoclinic symmetry onto monoclinic substrates when the growth surface is $(010)$ as well as all $(h0l)$ planes.\cite{https://doi.org/10.1002/pssb.201700134} Numerical results for the variation of the lattice parameters and the strain energy density were shown for hypothetical growth of $\beta$-(Al,Ga)$_2$O$_3$ and $\beta$-(In,Ga)$_2$O$_3$ on $\beta$-Ga$_2$O$_3$ for lattice mismatch induced biaxial strain. However, Grundmann assumed the stress free direction to be always aligned perpendicular to the surface. But this assumption is only valid during growth for highly symmetric crystal structures and growth planes with low crystallographic indices. Specifically, this assumption is only valid when the determinant of the stress tensor factorizes with the direction of growth separating from the reminder of the tensor. In this present work it is shown that the stress-free direction can take any direction in low-symmetry crystals. Hence, different stable unit cells emerge as possible stable solutions which can offer new opportunities in heteroepitaxial growth of low symmetry crystals.

\begin{figure}[ht]
\centering
\includegraphics[width=0.6\linewidth]{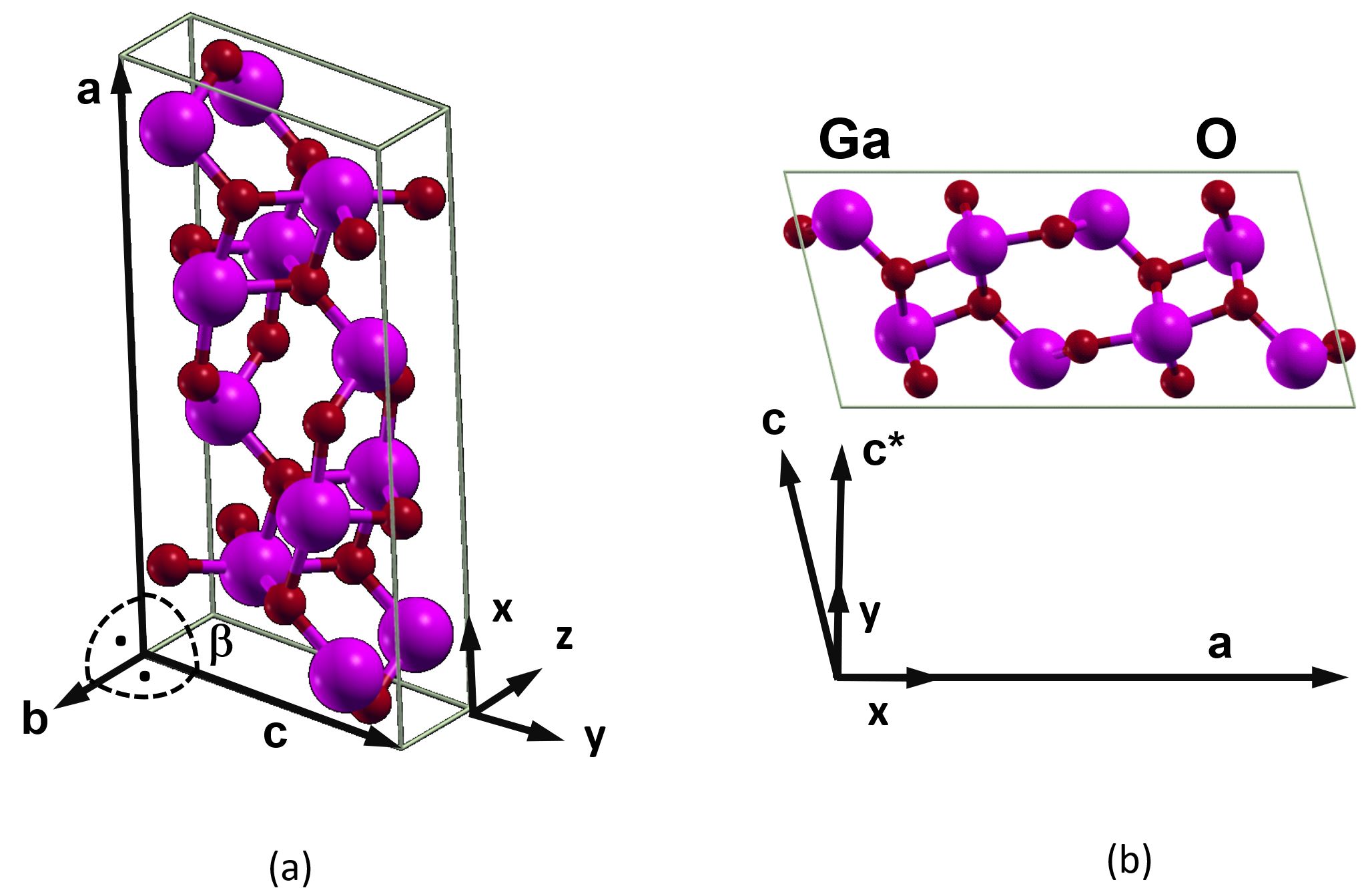}
\caption{(a) Unit cell of $\beta$-Ga$_2$O$_3$. Indicated are the monoclinic angle $\beta$ and the Cartesian coordinate system $(x,y,z)$ fixed to the unit
cell in this work with axes $\mathbf{a}$ and $\mathbf{b}$ tied to the $x$ and $-z$ directions, respectively. (b) View onto the $\mathbf{a}-\mathbf{c}$ plane along axis $\mathbf{b}$ which points into the plane. Indicated is the reciprocal lattice vector $\mathbf{c}^{\star}$ which is drawn not to scale. Reprinted from Ref.~\onlinecite{PhysRevB.93.125209} with copyright permission by American Physical
Society. }
\label{fig:unitcell}
\end{figure}

\begin{figure}[ht]
\centering
\includegraphics[width=1.0\linewidth]{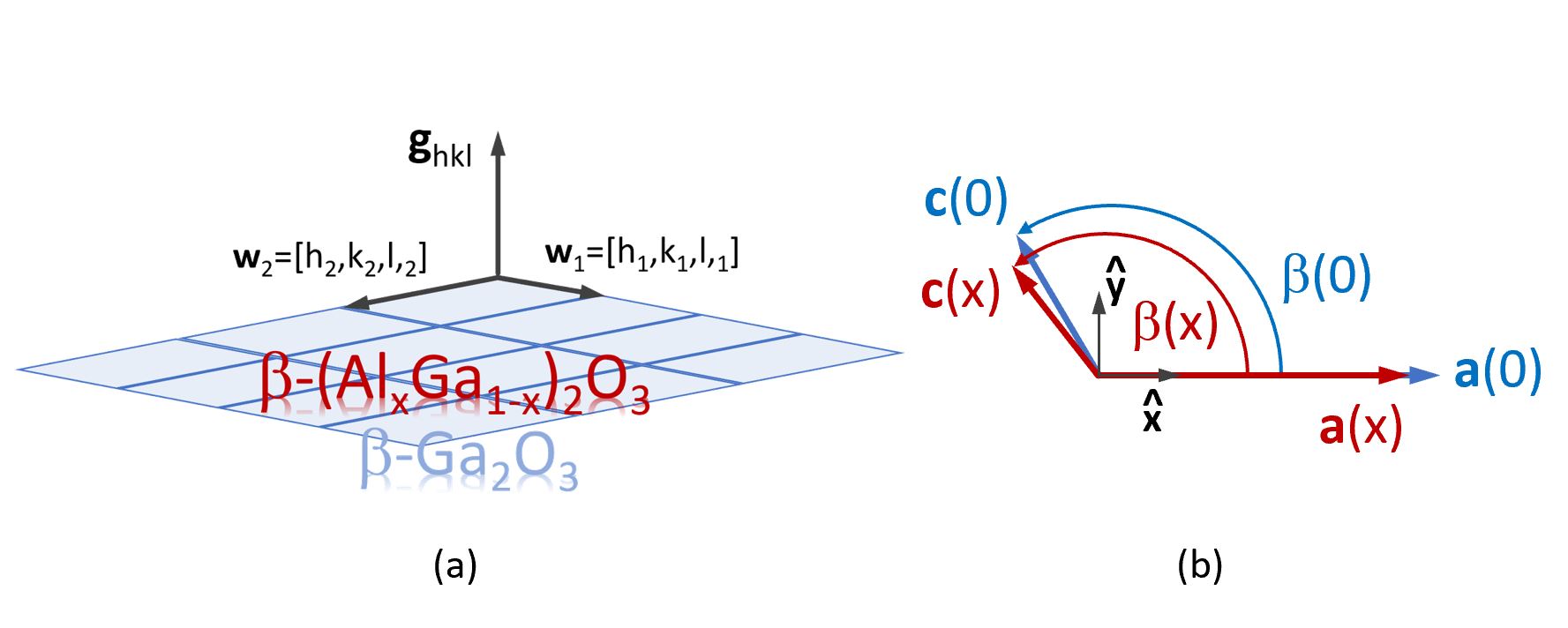}
\caption{(a) Definition of mesh vectors $\mathbf{w}_{1,2}$ parallel to the interface between the $\beta$-Ga$_2$O$_3$ substrate and the (Al$_x$Ga$_{1-x}$)$_2$O$_3$ epitaxial layer. Reciprocal plane vector $\mathbf{g}_{hkl}$is perpendicular to plane $(hkl)$ and pivotal in selecting $\mathbf{w}_{1,2}$. (b) View onto the monoclinic plane for the strain free unit cells of the substrate $(x=0)$ and the epitaxial layer $(x>0)$ with unit cell axes $\mathbf{a}(0)$, $\mathbf{c}(0)$ and $\mathbf{a}(x)$, $\mathbf{c}(x)$, respectively. A Cartesian coordinate system is selected within which the changes of the unit cell due to alloying is expressed, as shown. Note that the $z$ direction (out of the plane) is in opposite direction of $\mathbf{b}$ (not shown). For the $(h0l)$ strain discussed this work the unit cell retains its monoclinic symmetry.}
\label{fig:interfacemeshandunitcells}
\end{figure}

\section{Theory}
\subsection{The interface mesh vectors}
A given growth surface in a monoclinic lattice is characterized by unit cell vectors $\{\mathbf{a},\mathbf{b},\mathbf{c}\}$ and indexed using Miller indices for its crystallographic plane $(hkl)$. The surface normal vector parallels reciprocal vector $\mathbf{g}_{hkl}$
\begin{equation}
\mathbf{g}_{hkl}=h\mathbf{a}^{\star}+k\mathbf{b}^{\star}+l\mathbf{c}^{\star},
\end{equation}
\noindent with reciprocal lattice vectors defined as follows
\begin{equation}\label{eq:reciprocallatticevectordefinition}
\mathbf{v}_i\mathbf{v}^{\star}_j=\delta_{ij},
\end{equation}
\noindent where $\mathbf{v}_1 = \mathbf{a}$, $\mathbf{v}_2 = \mathbf{b}$, $\mathbf{v}_3 = \mathbf{c}$, $\mathbf{v}^{\star}_1 = \mathbf{a}^{\star}$, $\mathbf{v}^{\star}_2 = \mathbf{b}^{\star}$, $\mathbf{v}^{\star}_3 = \mathbf{c}^{\star}$, and $\delta_{ij}$ is the Kronecker symbol.

\subsection{Hooke's law}
The effects of stress and strain can be described by the introduction of the stress tensor $\sigma$ and a unit direction vector $\hat{\mathbf{e}}$ 
\begin{equation}
    \hat{\mathbf{e}}'-\hat{\mathbf{e}}=\sigma\mathbf{r},
\end{equation}
\noindent where $\sigma$ is a symmetric, rank-two tensor in units of kbar
\begin{equation}\label{eq:stresstensor}
\sigma=
\begin{pmatrix}
\sigma_{xx}& \sigma_{xy}& \sigma_{xz} \\
\sigma_{xy} & \sigma_{yy}& \sigma_{xy} \\
\sigma_{xz}& \sigma_{xy}& \sigma_{zz}
\end{pmatrix},
\end{equation}
\noindent and the magnitude of $\hat{\mathbf{e}}'-\hat{\mathbf{e}}$ is the amount of stress deviation from isotropic equilibrium in kbar under which the unit cell is in direction $\hat{\mathbf{e}}'$, and by the introduction of the strain tensor $\epsilon$ and lattice vector $\mathbf{w}$
\begin{equation}\label{eq:strainedvector}
\mathbf{w}'-\mathbf{w}=\epsilon\mathbf{w},
\end{equation}
\noindent a symmetric, dimensionless rank-two tensor
\begin{equation}\label{eq:straintensor}
    \epsilon=
\begin{pmatrix}
\epsilon_{xx}& \epsilon_{xy}& \epsilon_{xz} \\
\varepsilon_{xy} & \varepsilon_{yy}& \epsilon_{xy} \\
\epsilon_{xz}& \epsilon_{xy}& \epsilon_{zz}
\end{pmatrix},
\end{equation}
\noindent where the lattice vector under strain is denoted with a prime $'$. The generalized Hooke's law connects the strain tensor elements with the stress tensor elements
\begin{equation}
\tilde{\sigma}=C\tilde{\epsilon},   
\end{equation}
\noindent and
\begin{equation}\label{eq:linevectorsigma}
\tilde{\sigma}=\left(\sigma_{xx}, \sigma_{yy}, \sigma_{zz},\sigma_{xy}, \sigma_{xz}, \sigma_{yz}\right)^{T},
\end{equation}
\begin{equation}\label{eq:linevectorepsilon}
\tilde{\epsilon}=\left(\epsilon_{xx}, \epsilon_{yy}, \epsilon_{zz},2\epsilon_{xy}, 2\epsilon_{xz}, 2\epsilon_{yz}\right)^{T},
\end{equation}
\noindent where $T$ is the transpose operator, and the tensor of the elastic constants $C$ is given in units of kbar and in the Voigt notation with standard ordering.

\subsection{The epitaxial zero stress conditions}
During deposition of a crystalline overlayer on a crystalline substrate with a different lattice spacing than the epitaxial layer, the latter experiences stress. It is assumed here that the thickness of the epitaxial layer is negligible against that of the substrate. Hence, the substrate remains stress free, and the state of stress of the epitaxial layer is homogeneous and can be fully described by $\sigma$. The state of stress will depend on the amount of strain imposed onto the epitaxial layer by the mesh conditions across the interface as well as by the internal elastic forces within the epitaxial layer which cause further distortion of its unit cell until equilibrium is reached. When the elements of $C$ -- the elastic coefficients -- are known, and when the strain free unit cells of the substrate and the epitaxial layer are expressed in suitable coordinates and by known functional relationships, the strain and stress tensor elements for a given substrate surface $(hkl)$ and a given composition $x$ can be calculated.

The growing surface of the epitaxial layer is free to expand into the half space above the surface of the substrate. Hence, one can state that one or several direction(s) must exist, $\mathbf{v}$, along which the epitaxial layer is stress free, hence
\begin{equation}\label{eq:stressfreedirections}
0=\sigma\mathbf{v}.
\end{equation}
\noindent To begin with, these directions are not known. The necessary condition for any such direction to exist is the requirement that the determinant must vanish
\begin{equation}\label{eq:detsigmazero}
0=\det\sigma.
\end{equation}
\noindent A common paradigm states that stress free direction of an epitaxial layer is perpendicular to the surface. However, as shown in this work, this is only true for high symmetry orientations of which there are very few within the monoclinic crystal system of $\beta$-Ga$_2$O$_3$/$\beta$-(Al$_x$Ga$_{1-x}$)$_2$O$_3$.

\subsection{Vegard's rule}
The substitution of gallium by aluminum in $\beta$-(Al$_x$Ga$_{1-x}$)$_2$O$_3$ causes contraction of the length of all unit vectors and a small increase of the monoclinic angle. This is shown schematically in Fig.~\ref{fig:interfacemeshandunitcells}(b). A lattice vector $\mathbf{w}(x)$ may thus be written as 
\begin{equation}\label{eq:alloyedvector}
\mathbf{w}\left(x\right)=\mathbf{w}\left(x=0\right)+\delta\mathbf{w}\left(x\right),
\end{equation}
\noindent introducing vector $\delta\mathbf{w}\left(x\right)$. The explicit dependence of this vector as a function of composition must be known, or described, in order to establish the amount and type of strain a given epitaxial layer will experience towards a given template. For example, Vegard's rule may be used to express the behaviors of the magnitudes of the unit cell vectors, $a,b,c$, and angular orientation $ \beta$, as described by Kranert~\textit{et al.}\cite{doi:10.1063/1.4915627}
\begin{align}\label{eq:unitcellVegardrule}
    a\left(x\right)=&a\left(x=0\right)+x\delta a,\\
    b\left(x\right)=&b\left(x=0\right)+x\delta b,\\
    c\left(x\right)=&c\left(x=0\right)+x\delta c,\\    \beta\left(x\right)=&\beta\left(x=0\right)+x\delta \beta.
\end{align}
\noindent where $\delta a$, $\delta b$, $\delta c$, and $\delta \beta$ are parameters which render the linear change of the unit cell parameters with composition $x$. It is implicit that similar conditions exist for substitution of larger elements leading to expansion, or mixtures of expansion and contraction, and which is not further discussed. Instead, the case of isovalent Ga substitution by Al is considered here as example.

\subsection{The clamping conditions}
Figure~\ref{fig:interfacemeshandunitcells}(a) schematically depicts the mesh across a given interface characterized by crystallographic plane $(hkl)$. The mesh may be defined by two oblique vectors $\mathbf{w}_{1,2}$ parallel to the interface and perpendicular to reciprocal lattice vector $\mathbf{g}_{hkl}$
\begin{equation}
\mathbf{w}_j=h_j\mathbf{a}+k_j\mathbf{b}+l_j\mathbf{c},\mbox{} j=1,2.
\end{equation}
\noindent Miller indices $[h_j,k_j,l_j]$ maybe selected such that both vectors are within the surface and perpendicular to $\mathbf{g}_{hkl}$. In this respect, $\mathbf{g}_{hkl}$ is pivotal in selecting $\mathbf{w}_{j}$, however, the choices for the latter are infinite and the most convenient may be selected. 

Vectors $\mathbf{w}_{1,2}$ in Fig.~\ref{fig:interfacemeshandunitcells}(a) remain unchanged both in length and direction across the interface in the case of pseudomorphic growth. Hence,
\begin{equation}
\mathbf{w}_j\left(x=0\right)=\mathbf{w}'_j\left(x>0\right),\mbox{} j=1,2.
\end{equation}
\noindent The latter statement can be rewritten using Eq.~\ref{eq:strainedvector} and Eq.~\ref{eq:alloyedvector}
\begin{equation}
\mathbf{w}_j(0)=\left[\mathbf{w}_j(0)+\delta\mathbf{w}_j(x)\right]+\epsilon\mathbf{w}_j(x),\mbox{} j=1,2,
\end{equation}
\noindent which result in two general clamping conditions
\begin{equation}\label{eq:firstsecondclampingcondition}
0=\delta\mathbf{w}_j(x)+\epsilon\mathbf{w}_j(x),\mbox{} j=1,2.
\end{equation}
\noindent These conditions require that the changes in the strain free unit cell vectors due to alloying must equal their corresponding vectors' negative change under the pseudomorphic strain. Note that Eqs.~\ref{eq:stressfreedirections}, ~\ref{eq:detsigmazero}, and~\ref{eq:firstsecondclampingcondition} are stated within the Cartesian coordinate system of the strain free unit cell for the epitaxial layer. The elements of $\epsilon$ bring lattice vectors $\mathbf{w}_j(x)$ to coincide with $\mathbf{w}_j(0)$, in length and direction. The axes of deformation expressed in the tensor elements may not coincide with any of the in-plane or out-of-plane directions of a given surface, instead, the elements of $\epsilon$ will adjust to whichever way a given unit cell is deformed under the clamping condition. This definition of the elements of $\epsilon$ here is useful because the tensor elements refer to the changes of the intrinsic lattice parameters, $a$, $b$, $c$, and $\beta$ for a given epitaxial layer, regardless of the actual orientation of the unit cell within a given epitaxial layer relative to the substrate. In this respect, the coordinate system for the substrate is irrelevant.

\subsection{The strain-stress relationship for $(h0l)$ interface clamping}
The approach described above is demonstrated for heteroepitaxial systems with monoclinic crystal symmetry in the present work. The cases are reduced to the situations where axis $\mathbf{b}$ is parallel to the surface, i.e., the class of surfaces characterized by $(h0l)$ is discussed here. A generalization of this approach to triclinic cases which includes the situation when monoclinic epitaxial layers are distorted to triclinic symmetry is subject of a different work.

Surface vectors, $\mathbf{w}_1=[h_1,k_1,l_1]$ and $\mathbf{w}_2=[h_2,k_2,l_2]$, are selected to be perpendicular to each other and to the surface normal
\begin{equation}
\mathbf{w}_1=h_1\mathbf{a}+l_1\mathbf{c}=[l0-h],
\end{equation}
\begin{equation}
\mathbf{w}_2=k_2\mathbf{b}=[010],
\end{equation}
\begin{equation}
\mathbf{g}_{h0l}=h\mathbf{a}^{\star}+l\mathbf{c}^{\star},
\end{equation}

\noindent and $k_1=0,h_2=0,l_2=0$. The selection $h_1=l$ and $l_1=-h$ places $\mathbf{w}_1$ parallel to the surface, and the scalar products vanish $\mathbf{w}_1\mathbf{w}_2=0$, $\mathbf{w}_1\mathbf{g}_{h0l}=0$, and  $\mathbf{w}_2\mathbf{g}_{h0l}=0$.

The strain free lattice variation vectors $\delta\mathbf{w}_j$ are obtained from the difference between the strain free vectors within the alloy for $x>0$ and $x=0$
\begin{align}\label{eq:strainfreelatticevariationmonoclinic}
\mathbf{a}(0)=&a(0)\hat{x},\\\label{eq:strainfreelatticevariationmonoclinic1}
\mathbf{c}(0)=&-c(0)\cos\left[\pi-\beta(0)\right]\hat{x}+c(0)\sin\left[\pi-\beta(0)\right]\hat{y},\\\label{eq:strainfreelatticevariationmonoclinic2}   
\mathbf{a}(x)=&a(x)\hat{x},\\\label{eq:strainfreelatticevariationmonoclinic3}
\mathbf{c}(x)=&-c(x)\cos\left[\pi-\beta(x)\right]\hat{x}+c(x)\sin\left[\pi-\beta(x)\right]\hat{y}, 
\end{align}
\begin{equation}
\delta\mathbf{a}(x)=\mathbf{a}(x)-\mathbf{a}(0),
\end{equation}
\begin{equation}
\delta\mathbf{c}(x)=\mathbf{c}(x)-\mathbf{c}(0),
\end{equation}
\noindent where $\hat{x}$ and $\hat{y}$ are unit vectors along Cartesian directions $x$ and $y$ within the monoclinic plane of the strain free epitaxial layer (Fig.~\ref{fig:interfacemeshandunitcells}(b)).

The clamping conditions are
\begin{equation}\label{eq:clamph0l1}
    0=\left[l\mathbf{a}(x)-h\mathbf{c}(x)\right]\epsilon+\left[l\delta\mathbf{a}(x)-h\delta\mathbf{c}(x)\right],
\end{equation}
\begin{equation}\label{eq:clamph0l2}
    0=\mathbf{b}(x)\epsilon+\delta\mathbf{b}(x),
\end{equation}
\begin{equation}\label{eq:clamph0l3}
    0=\det\sigma.
\end{equation}

Equation~\ref{eq:clamph0l2} permits to determine $\epsilon_{zz}$
\begin{equation}
    \epsilon_{zz}=\frac{b(0)-b(x)}{b(x)}.
\end{equation}

\noindent Equations~\ref{eq:clamph0l1} and~\ref{eq:clamph0l3} must be combined. The first clamping statement can be written into two equations

\begin{widetext}\label{eq:h0lexamplestrainclamping1}
\begin{align}\label{eq:lhequation1}
    0=-\delta c_x h x + \delta a_x l x - \epsilon_{xy} h (c_y + \delta c_y x) + \epsilon_{xx} (l (a_x + \delta a_x x) - h (c_x + \delta c_x x)),
\end{align}
\begin{align}\label{eq:lhequation2}
    0=-\delta c_y h x - \epsilon_{yy} h (c_y + \delta c_y x) + \epsilon_{xy} (l (a_x + \delta a_x x) - h (c_x + \delta c_x x)).
\end{align}
\end{widetext}
\noindent This set of equations is under-determined. Assuming, for example, $\epsilon_{yy}$ would be known, then one can find closed expressions for $\epsilon_{xx}$ and $\epsilon_{xy}$
\begin{widetext}
\begin{align}\label{eq:finalseth0l1}
\epsilon_{xx}=&\frac{c_y^2 \epsilon_{yy} h^2 + c_y \delta c_y (1 + 2 \epsilon_{yy}) h^2 x + x (a_x l (\delta c_x h - \delta a_x l) + c_x h (-\delta c_x h + \delta a_x l)}{(c_x h - a_x l + \delta c_x h x - \delta a_x l x)^2}\\
&+ \frac{(-\delta c_x^2 h^2 + \delta c_y^2 (1 + \epsilon_{yy}) h^2 + 2 \delta a_x \delta c_x h l - \delta a_x^2 l^2) x)}{(c_x h - a_x l + \delta c_x h x - \delta a_x l x)^2},\\
\epsilon_{xy}=&\frac{h (c_y \epsilon_{yy} + \delta c_y (1 + \epsilon_{yy}) x)}{-c_x h + a_x l - \delta c_x h x + \delta a_x l x}.
\end{align}
\end{widetext}
\noindent The elements of $\sigma$ can be expressed via the elastic tensor elements and $\epsilon_{xx},\epsilon_{xy},\epsilon_{yy}$ ($ij=xx,xy,yy$)
\begin{align}\label{eq:h0lexamplestrainclamping2}
\sigma_{ij}=C_{ij,xx}\epsilon_{xx}+C_{ij,yy}\epsilon_{yy}+C_{ij,zz}\epsilon_{zz}+C_{ij,xy}2\epsilon_{xy}.
\end{align}
\noindent Equation~\ref{eq:clamph0l3} requires the determinant of $\sigma$ to vanish, and which simplifies for the monoclinic symmetry unit cell
\begin{equation}\label{eq:finalseth0l2}
0=\sigma_{zz}\sigma_{xx}\sigma_{yy}-\sigma_{zz}\sigma^2_{xy}.
\end{equation}
\noindent Inserting the expressions for $\epsilon_{xx}$ and $\epsilon_{xy}$ above into the ``$\det\sigma=0$'' condition leaves one equation left which is of third order in the remaining coefficient $\epsilon_{yy}$. Hence, in general, the ``$\det\sigma=0$'' condition leads to three possible solutions. Hence, three unit cells exists which are stable from geometrical and elastic property perspectives and which match the requirement to fit the mesh defined by vectors $\mathbf{w}_{1,2}$. Each solution renders a different strain tensor, and subsequently, differ in unit lattice vectors, stress tensor elements, stress-free direction, unit cell volume, total elastic energy volume density, and lattice spacing along growth direction $\mathbf{g}_{h0l}$, $d_{h0l}$. In the following, these different solutions will be discussed and shown for the example of a hypothetical epitaxial layer with 10$\%$ aluminum content grown pseudomorphic on $\beta$-Ga$_2$O$_3$ with $(h0l)$ surfaces.\footnote{Note the structure of the determinant for the stress tensor where component $\sigma_{zz}$ always factorizes out. This has trivial solutions then for all cases when the $b$ axis is perpendicular to the surface, (010). Then the $z$ direction is also parallel to the surface normal $\mathbf{g}_{hkl}$. Hence, for growth on $(010)$ only one stress free solution with a stress free direction not parallel to the surface exists which is trivially then the surface normal. For almost all other crystallographic surfaces, the stress free directions are neither parallel nor perpendicular to the surface and all represent potential stable unit cells during growth.}

\subsection{Definition of the rotation system to address all $(h0l)$}
\begin{figure}[ht]
\centering
\includegraphics[width=0.7\linewidth]{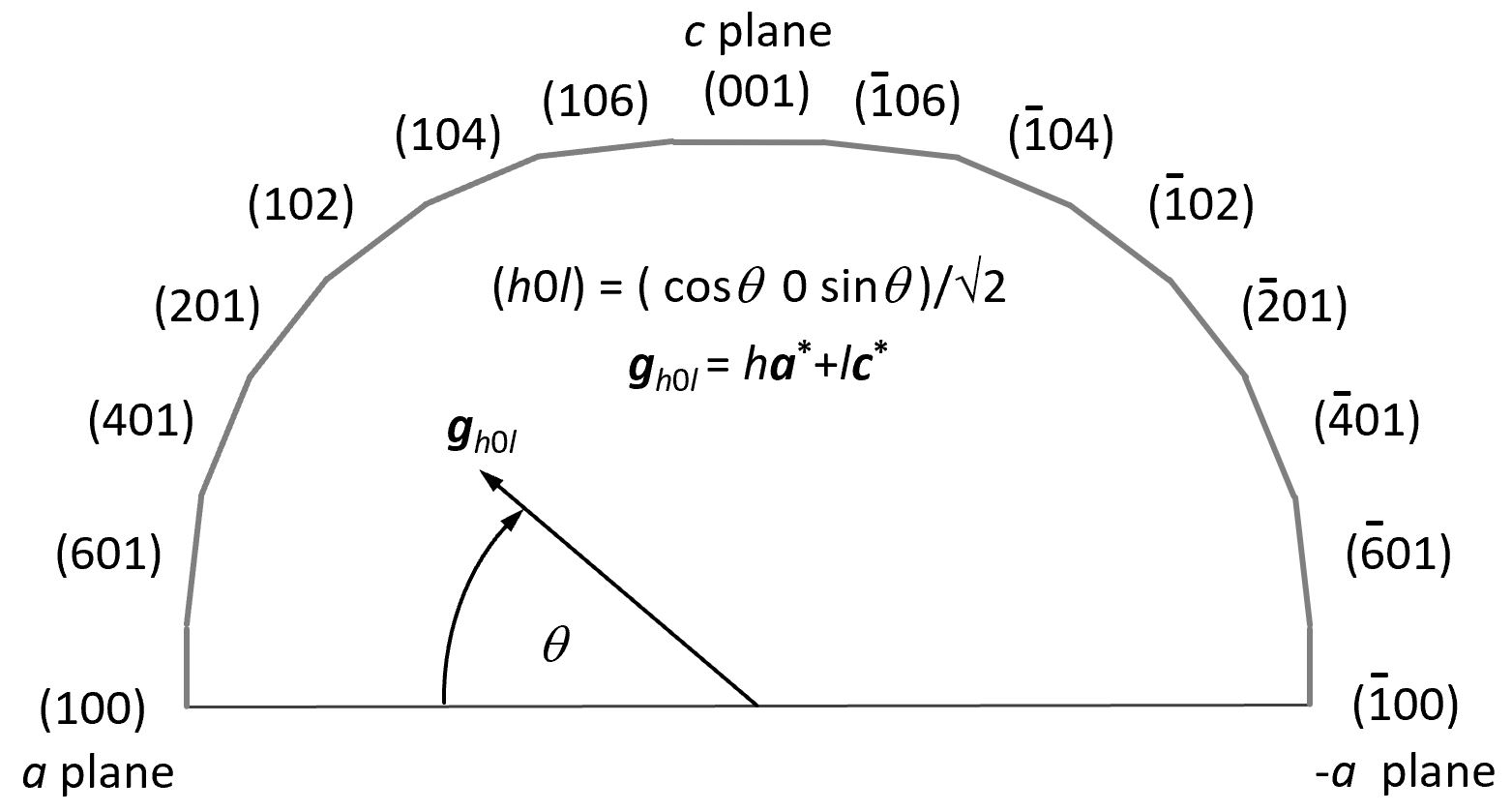}
\caption{Definition of rotation $\theta$ used here to parameterize the Miller indices in surface plane $(h0l)$.}
\label{fig:h0lthetaplane}
\end{figure}
To address a continuous variation of the template surface plane, parameter $\theta$ is introduced such that
\begin{equation}
(h0l)\rightarrow(\cos\theta 0 \sin\theta)/\sqrt{2}.
\end{equation}
\noindent Note that a factor common to both $h$ and $l$ can be divided out from Eqs.~\ref{eq:lhequation1} and ~\ref{eq:lhequation2}, hence, $\theta=0,\pi/2,\pi$. etc. is equivalent to $(100), (010), (\bar{1}00)$. etc., respectively.

\section{Parameter details}
\subsection{Elastic coefficients}
To begin with, we provide the numerical values for the elastic coefficients for the strain free alloy in Eq.~\ref{eq:C_AlGO} in units of kbar/(unit strain), which are obtained from Ref.~\onlinecite{PhysRevApplied.18.064019}.
\begin{widetext}
\begin{equation}
\label{eq:C_AlGO}
C_{(Al_{x}Ga_{1-x})_2O_3} = 
\begin{bmatrix}
 2143+505x & 1103+10x & 1200+90x & 0 & -197-76x & 0 \\
 1103+10x & 3300+546x & 669-80x & 0 & 119+24x & 0 \\
 1200+90x & 669-80x & 3248+748x & 0 & 75+71x & 0 \\
 0 & 0 & 0 & 500+197x & 0 & 182+9x \\
 -197-76x & 119+24x & 75+71x & 0 & 689+329x & 0 \\
 0 & 0 & 0 & 182+9x & 0 & 949+271x
\end{bmatrix}.
\end{equation}
\end{widetext}

\subsection{Lattice parameters}
\noindent For calculation of the strain free lattice parameters we use the numerical results provided by Kranert~\textit{et al.} in Ref.~\onlinecite{doi:10.1063/1.4915627}, with $a(x=0) = 12.30924867 \AA$, $\delta a =-0.42\AA$, $b(x=0) = 3.05596658\AA$, $\delta b = - 0.13\AA$, $c(x=0) = 5.82578805\AA$, $\delta c = - 0.17\AA$, $\beta(x=0) = 103.718941556756^{\circ}$, and $\delta \beta = 0.31^{\circ}$. Then, Eqs.~\ref{eq:finalseth0l1}--\ref{eq:finalseth0l2} can be solved either analytically or numerically.

\section{Results and Discussions}
\subsection{Calculation of the strain tensor solutions}
\begin{figure}[ht]
\centering
\includegraphics[width=1.0\linewidth]{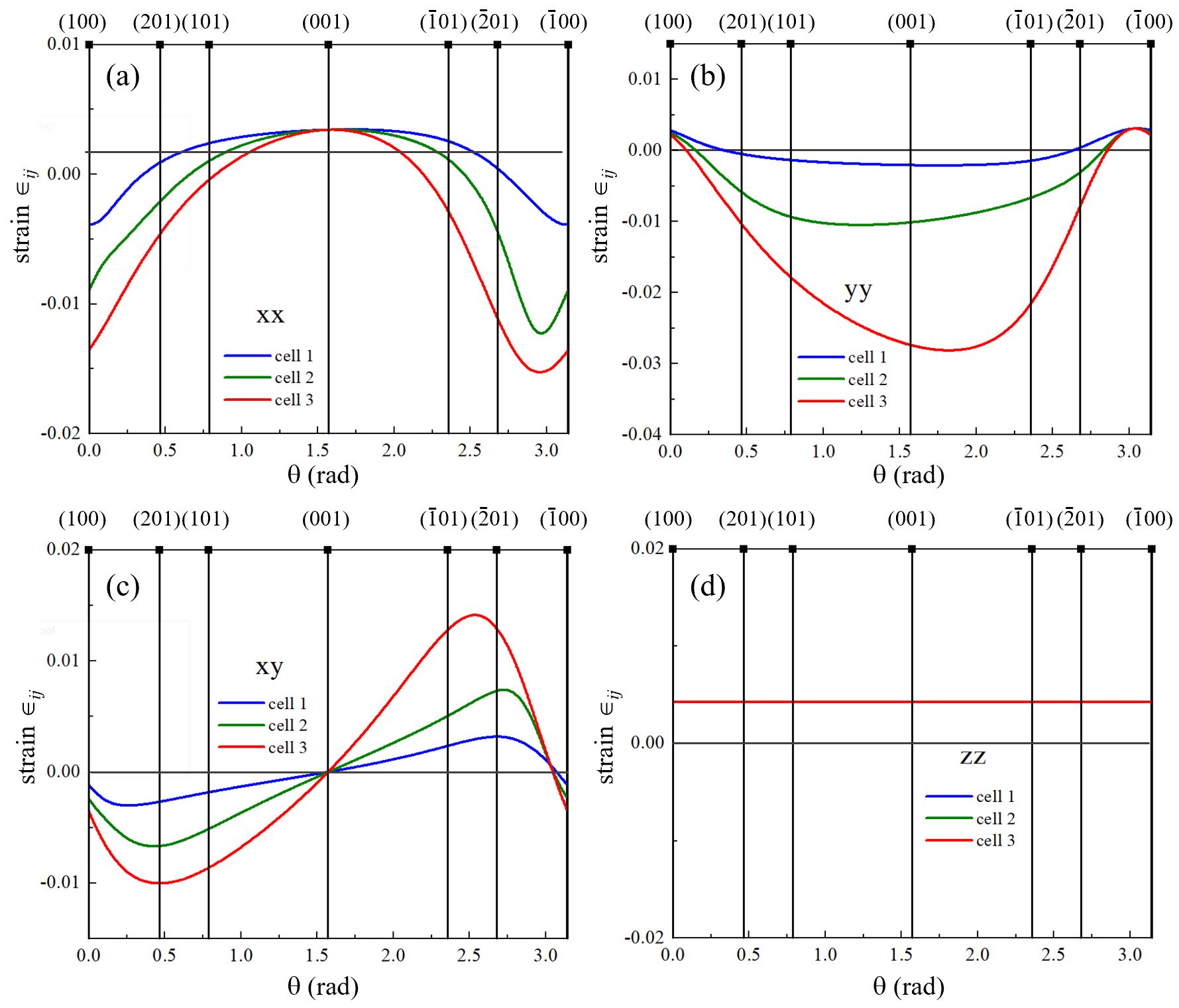}
\caption{Strain elements $\epsilon_{xx}$, $\epsilon_{xy}$, $\epsilon_{yy}$, and $\epsilon_{zz}$ for a fully strained $\beta$-(Al$_{0.1}$Ga$_{0.9}$)$_2$O$_3$ epitaxial layer clamped onto a $\beta$-Ga$_2$O$_3$ template with $(h0l)$ surface. Rotation $\theta$ is indicated in Fig.~\ref{fig:h0lthetaplane}. Three differently strained unit cell solutions are found here (cell 1: blue; cell 2: olive; cell 3: red).}
\label{fig:h0lstrainx10bGO}
\end{figure}
Figure~\ref{fig:h0lstrainx10bGO} depicts the four strain elements, $\epsilon_{xx}$, $\epsilon_{xy}$, $\epsilon_{yy}$, and $\epsilon_{zz}$ for the three unit cell solutions for a fully strained  $\beta$-(Al$_{0.1}$Ga$_{0.9}$)$_2$O$_3$ epitaxial layer clamped onto a $\beta$-Ga$_2$O$_3$ template with $(h0l)$ surface. The abscissas are the rotation $\theta$ indicated in Fig.~\ref{fig:h0lthetaplane}. Groups of three plots are shown for three differently strained unit cells found from the solution described above. The three solutions are color coded (cell 1: blue; cell 2: olive; cell 3: red) and the same color code is used in all subsequent figures unless otherwise noted. The element $\epsilon_{zz}$ is constant versus rotation, and equal for all three cells. This is because the $\mathbf{b}$ axis of the epitaxial layer is clamped onto the $\mathbf{b}$ axis of the substrate, and does not change regardless of the Miller indices $h$ and $l$. Furthermore, in Fig.~\ref{fig:h0lstrainx10bGO}(a) the strain $\epsilon_{xx}$ vanishes for all cells for the $c$-plane because there the $\mathbf{a}$ axis of the epitaxial layer is clamped onto the $\mathbf{a}$ axis of the substrate. Likewise, in Fig.~\ref{fig:h0lstrainx10bGO}(b) the strain $\epsilon_{yy}$ vanishes for all cells for the $a$-plane and $-a$-plane because there the $(-)\mathbf{c}$ axis of the epitaxial layer is clamped onto the $(-)\mathbf{c}$ axis of the substrate. The shear strains cross zero for the $c$-plane surface, and at another high indexed plane close to the $a$-planes.

\subsection{Calculation of the stess tensor solutions}
\begin{figure}[ht]
\centering
\includegraphics[width=1.0\linewidth]{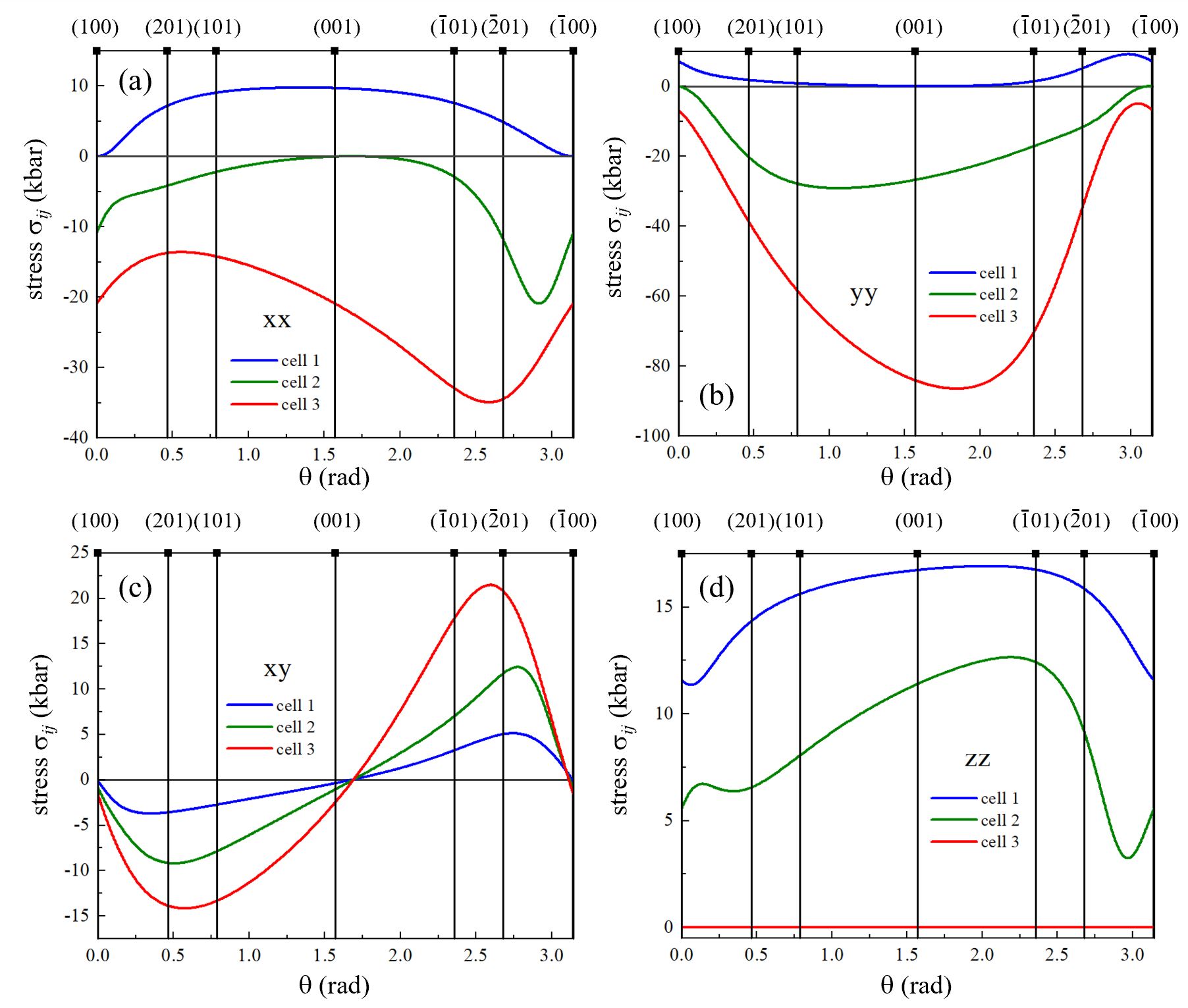}
\caption{Same as Fig.~\ref{fig:h0lstrainx10bGO} for the stress tensor elements.}
\label{fig:h0lstressx10bGO}
\end{figure}
Figure~\ref{fig:h0lstressx10bGO} depicts the four stress elements, $\sigma_{xx}$, $\sigma_{xy}$, $\sigma_{yy}$, and $\sigma_{zz}$ for the same set of strained unit cell solutions shown in Fig.~\ref{fig:h0lstrainx10bGO}. The element $\sigma_{zz}$ is zero regardless of rotation for the third unit cell. It is noteworthy to mention that except for $\sigma_{xy}$ there is no common crossing point. It is further of interest to note that unit cell 1 has zero stress in $\sigma_{xx}$ for $(100)$ and $(\bar{1}00)$ as well as in $\sigma_{yy}$ for $(010)$, while unit cell 2 has zero stress in $\sigma_{yy}$ for $(100)$ and $(\bar{1}00)$ as well as in $\sigma_{xx}$ for $(010)$. It is obvious that both cells appear to be complementary of some kind, and which will become clearer soon. It is also noteworthy to observe that the shear elements have common zero crossing points at two different, high indexed Miller planes.

\subsection{Unit cell parameters solutions}
\begin{figure}[ht]
\centering
\includegraphics[width=0.9\linewidth]{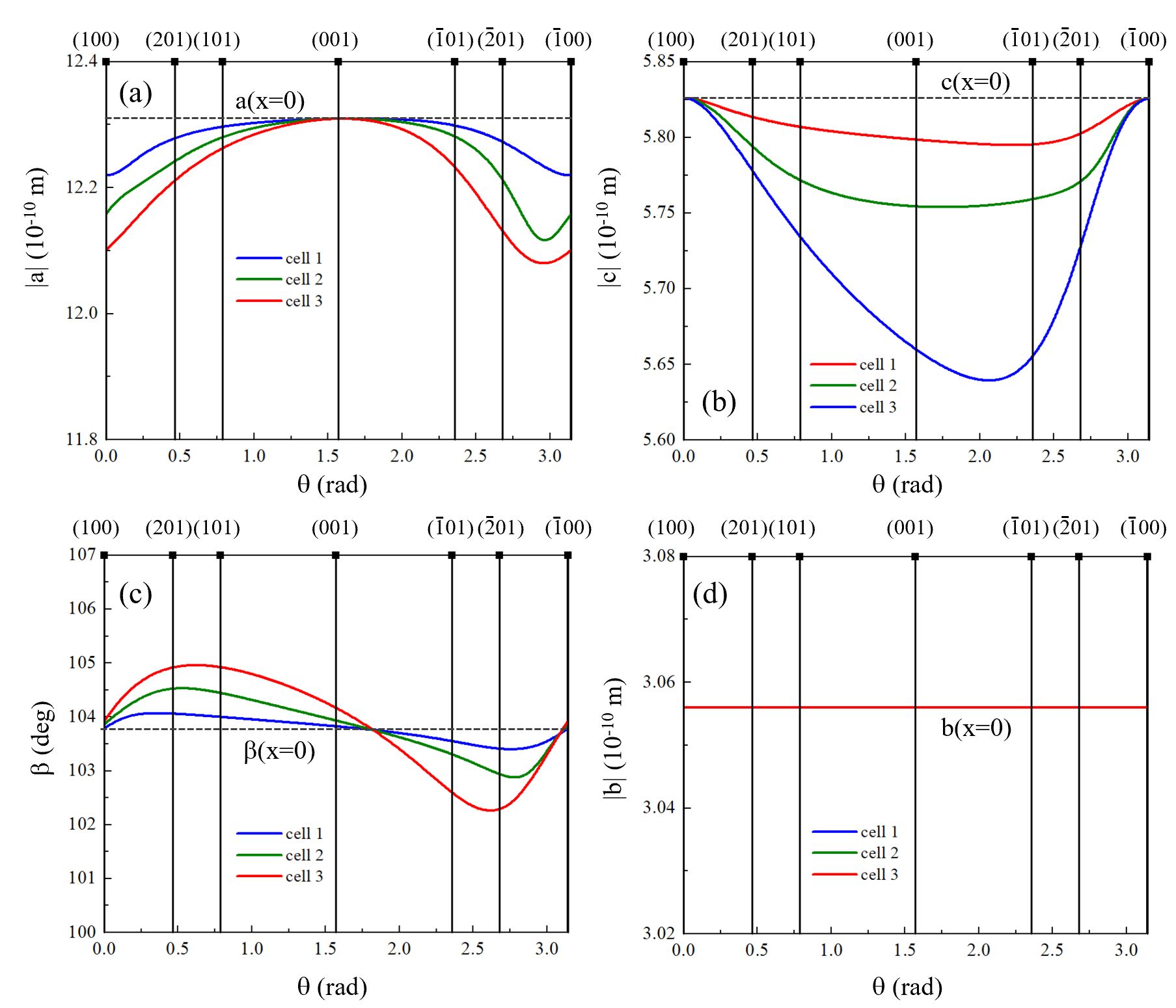}
\caption{Same as Fig.~\ref{fig:h0lstrainx10bGO} for the magnitudes of the epitaxial layer unit cell lattice vectors ($|\mathbf{a}|$: (a), $|\mathbf{c}|$: (b), $|\mathbf{b}|$: (d)) and monoclinic angle $\beta$ (c). Values for the strain free $\beta$-Ga$_2$O$_3$ substrate are indicated by dashed horizontal lines. }
\label{fig:h0lstrainx10bGOvectors}
\end{figure}
Figure~\ref{fig:h0lstrainx10bGOvectors} depicts the magnitudes of the unit cell lattice vectors and monoclinic angle $\beta$ for the three differently strained unit cells. In accordance with the strain elements, $|\mathbf{a}|$ is common to all unit cells for the $c$-plane, and $|\mathbf{c}|$ for $(100)$ and $(\bar{1}00)$. $|\mathbf{b}|$ is constant for all cells and regardless of surface orientation $h$, $l$. The variations of the lattice vector lengths reflect the possible changes among the four lattice parameters to adapt to the respective template by maintaining the clamping conditions and by satisfying the stress free condition.  

\subsection{Unit cell volume, $d_{hkl}$, elastic unit cell energy, and stress free directions}
\begin{figure}[ht]
\centering
\includegraphics[width=1.0\linewidth]{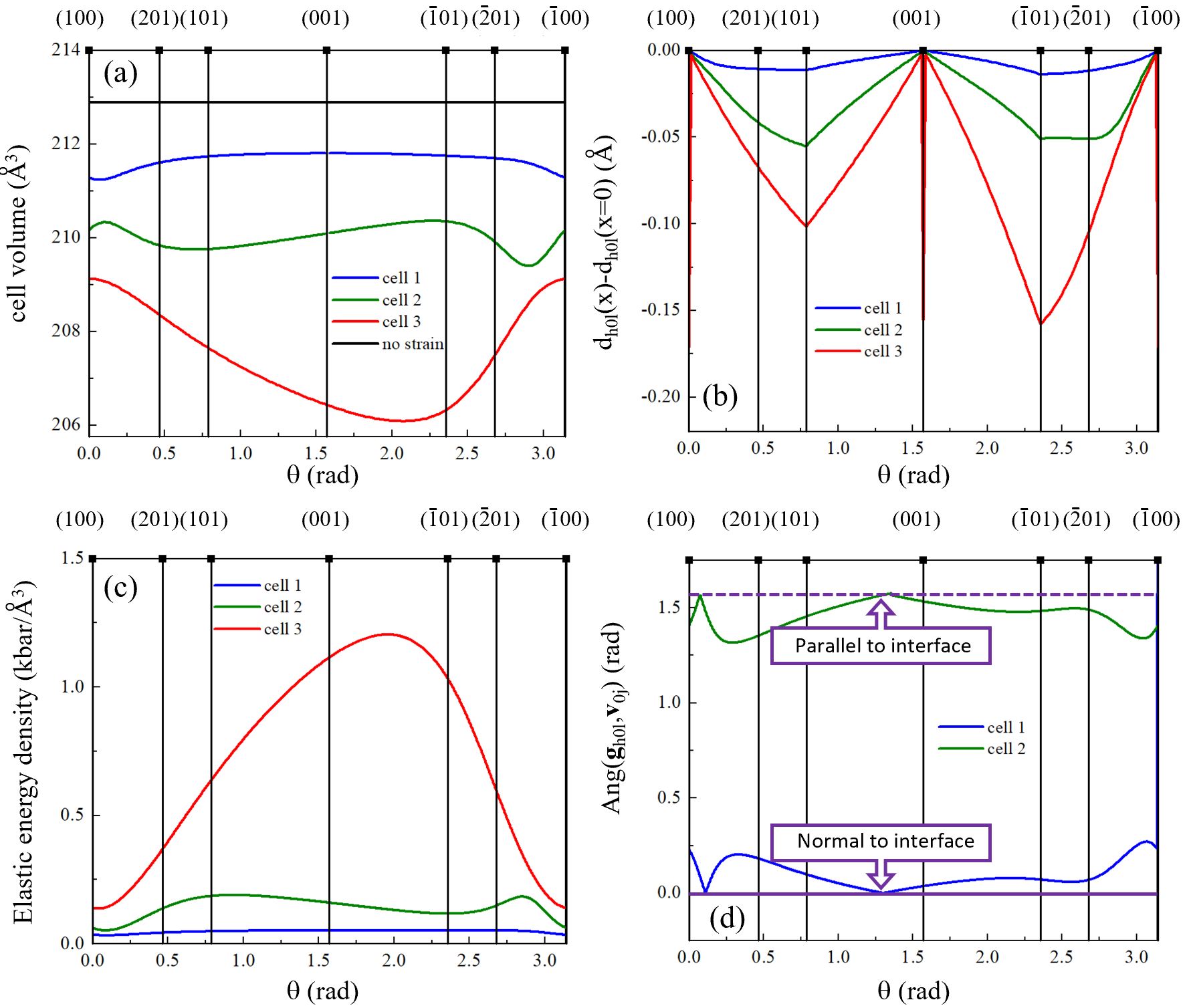}
\caption{Same as Fig.~\ref{fig:h0lstrainx10bGO} for the elastic energy (volume) density (a), the lattice spacing in growth direction, $d_{h0l}$ (b), the unit cell volume (c), and the stress free directions (d). Note that the third unit cell (red) stress free direction is parallel along axis $\mathbf{b}$ regardless of rotation $\theta$ and not shown in (d). The stress free directions are indicated by the angular difference between the stress free direction in the Cartesian coordinate system of the unit cell in Fig.~\ref{fig:interfacemeshandunitcells}(b) and the rotation $\theta$. Thereby, when the difference is zero, the stress free direction point along the surface normal, when the difference is $\pm \pi/2$ the direction is exactly within the surface. As can be seen, for most of the surface Miller indices the stress free directions point obliquely to the surface where cell 1 and cell 2 are possible competing domains during growth.}
\label{fig:h0lstrainx10voldh0lzerostress}
\end{figure}
To this end, a differentiation is necessary to identify which if the three cells are most, more, or not likely at all to develop during epitaxial growth. Figure~\ref{fig:h0lstrainx10voldh0lzerostress}(a) shows the volume of the three differently strained unit cells versus rotation, together with the volume of the strain free unit cell of $\beta$-(Al$_{0.1}$Ga$_{0.9}$)$_2$O$_3$. The unit cell volume is calculated by 
\begin{equation}
V=\mathbf{a}\left(\mathbf{b}\times\mathbf{c}\right),
\end{equation}
\noindent where $\times$ is the cross product. In accordance with the behavior of strain elements which are overall largest for cell 3 and smallest for cell 1, the unit cell volume is largest for the least strained cell 1 and smallest for the strongest strained cell 3. The same property is reflected by the behavior of the lattice plane distances along the surface normal, $d_{h0l}$
\begin{equation}
d_{h0l} = 
 \frac{\sin\beta}{\sqrt{\frac{h^2}{a^2} + \frac{k^2\sin^2\beta}{b^2} + \frac{l^2}{c^2} - \frac{2hl\cos\beta}{ac}}}.
\end{equation}
\noindent Figure~\ref{fig:h0lstrainx10voldh0lzerostress}(b) shows the differences of lattice spacing for the three differently strained lattices of the epitaxial layer versus rotation relative to $d_{h0l}$ of the strain free lattice of $\beta$-(Al$_{0.1}$Ga$_{0.9}$)$_2$O$_3$. The lattice spacing is identical for the directions with lowest Miller indices, $(100)$, $(010)$ and $(\bar{1}00)$, and largest for planes $(101)$ and $(\bar{1}01)$. The cell with largest strain and smallest volume reflects also the largest changes in lattice spacing overall. The amount of elastic potential energy stored within the strained unit cells per unit displacement can be calculated from the strain and stress tensor elements
\begin{equation}
u=\frac{1}{2}\sum_{i=1}^{6}\tilde{\epsilon}_{i}\tilde{\sigma}_{i},
\end{equation}
\begin{equation}
u=\frac{1}{2}\sum_{i=1}^{6}\sum_{j=1}^{3}C_{ij}\tilde{\epsilon}_{i}\tilde{\epsilon}_{j},
\end{equation}
\noindent where
\begin{equation}
\tilde{\sigma}=\left(\sigma_{1}, \sigma_{2}, \sigma_{3},\sigma_{4}, 0, 0\right)^{T},
\end{equation}
\begin{equation}
\tilde{\epsilon}=\left(\epsilon_{1}, \epsilon_{2}, \epsilon_{3},\epsilon_{4}, 0, 0\right)^{T},
\end{equation}
\noindent and $\tilde{\sigma}_i$ and $\tilde{\epsilon}_i$ are defined in Eqs.~\ref{eq:linevectorsigma} and~\ref{eq:linevectorepsilon}, respectively. Figure~\ref{fig:h0lstrainx10voldh0lzerostress}(c) shows $u$ for the three differently strained lattices of the epitaxial layer versus rotation. As expected, cell 1 with lowest strain overall reflects the lowest elastic potential energy, while cell 3 reveals very large potential energy, It is thus clear that cell 1 is most likely to form during epitaxial growth while cells 2 and 3 will require much higher energy to form. However, in addition to the elastic potential energy it is also important to evaluate the stress free directions for every unit cell solution. The requirement that the determinant vanishes for the stress tensor was only the necessary condition for a stress free cell to exist. The final question to answer is in which direction the zero stress within a given unit cell is pointing. Since the determinant of $\sigma$ vanishes for all 3 unit cells, for each cell there must be at least one direction $\mathbf{v}_{0,j}$ along which the stress is zero
\begin{equation}
\sigma\mathbf{v}_{0,j}=0, j=1,2,3.
\end{equation}
Inspection of Fig.~\ref{fig:h0lstressx10bGO}(d) reveals that for the unit cell 3 element $\sigma_{zz}$ is zero throughout the rotation. Figure~\ref{fig:h0lstrainx10bGOvectors}(d) reveals that the length of axis $\mathbf{b}$ is constant for unit cell 3 as well. Hence, $\mathbf{v}_{0,3}$ is parallel to direction $z$, that is, parallel to the surface, regardless of $h$ and $l$. Epitaxial growth of this cell is not possible because no stress can be exerted onto the epitaxial layer from direction perpendicular to the template surface. The stress free direction being parallel to $z$ in this scenario means that one can deform the unit cell of the epitaxial layer along $\mathbf{w}_1$ and along the surface normal until the internal elastic forces contract the length along axis $\mathbf{b}$ such that the latter matches the axis of the substrate. The stress free directions for cell 1 and cell 2 can be calculated from the data shown in Figs.~\ref{fig:h0lstressx10bGO}. The vectors $\mathbf{v}_{0,j}, j=1,2$ are
\begin{equation}
\mathbf{v}_{0,j}=-\sigma_{xx,j}\hat{x}+\sigma_{xy,j}\hat{y}=-\sigma_{xy,j}\hat{x}+\sigma_{yy,j}\hat{y},
\end{equation}
\noindent where $\sigma_{\dots,j}$ indicated elements of the stress tensor of the strained unit cell $j=1,2$. Hence, vectors $\mathbf{v}_{0,j}$ are expressed within the unit cell of the strained epitaxial layer and within the coordinates shown in Fig.~\ref{fig:interfacemeshandunitcells}. The angle between the stress free direction and the surface vector, $\mathbf{g}_{h0l}$, $\cos^{-1}\left(\frac{\mathbf{g}_{h0l}\mathbf{v}_{0,j}}{g_{h0l}v_{0,j}}\right)$ is plotted in Fig.~\ref{fig:interfacemeshandunitcells}(d). Note that vector $\mathbf{g}_{h0l}$ is different for every of the strained unit cell solutions discussed here since vectors $\mathbf{a}$ and $\mathbf{c}$ differ for each cell as shown in Fig.~\ref{fig:h0lstrainx10bGOvectors}. However, the epitaxial layer is lined up with the template such that $\mathbf{g}_{h0l}$ between substrate and layer are parallel to each other. This is ensured because of the clamping conditions which require that vectors $\mathbf{w}_{1,2}$ line up between the substrate and the epitaxial layer, and which do so by changing vectors $\mathbf{a}$ and $\mathbf{c}$ via the respective strain tensors for each unit cell solution. Hence, the $\mathbf{g}_{h0l}$ in every cell also renders the surface normal for the epitaxial layer. As can be seen in  Fig.~\ref{fig:interfacemeshandunitcells}(d), cell 1 has its stress free direction close to the surface normal while that of cell 2 is mostly around the surface itself. However, except for two specific surfaces, both directions deviate from the normal and parallel. First, therefore, the stress free direction is not always perpendicular to the surface, in fact, it is not for most of the $(h0l)$ planes discussed here, second, the second cell solution has its stress free direction also not parallel to the surface. Hence, the second cell could also grow albeit requiring higher densification due to the reduced lattice spacing and the smaller unit cell parameters, and therefore requiring higher elastic potential deformation energy. Nonetheless, during non-equilibrium growth such as in metal organic vapor phase epitaxy (MOVPE) or molecular beam epitaxy (MBE) formation of competing domains with different unit cell dimensions is possible.  

\subsection{Band-to-band transitions in pseudomorphically strained (Al$_{0.1}$Ga$_{0.9}$)$_2$O$_3$/$(h0l)$Ga$_2$O$_3$}
\begin{figure}[ht]
\centering
\includegraphics[width=0.9\linewidth]{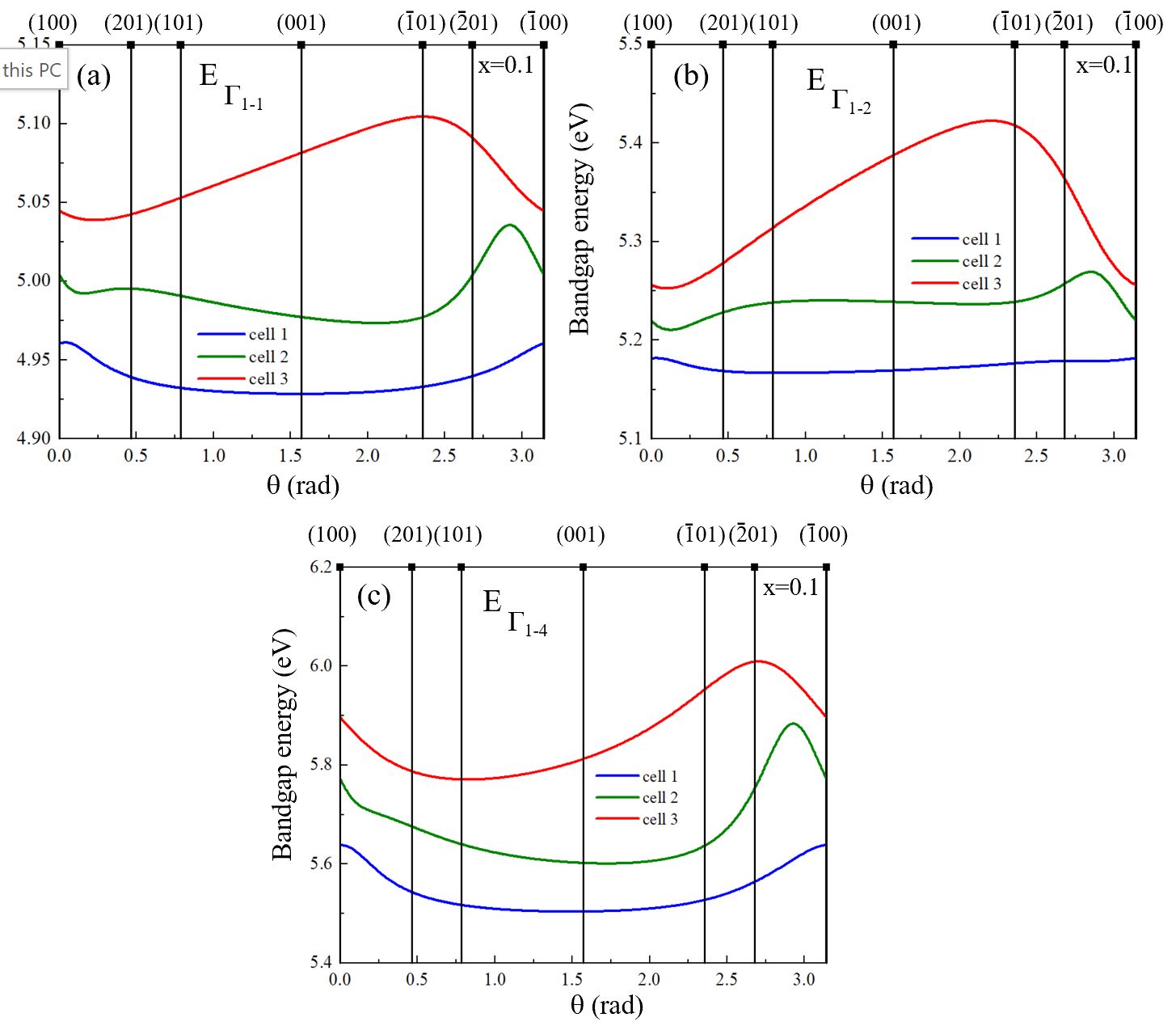}
\caption{Same as Fig.~\ref{fig:h0lstrainx10bGO} for the band-to-band transitions in the strained hypothetical $\beta$-(Al$_{0.1}$Ga$_{0.9}$)$_2$O$_3$ epitaxial layer. (a) transition energy between $\Gamma$-point conduction band level $c-1$ and valence band level $v-1$, (b) $c-1$-$v-2$, (c) $c-1$-$v-4$. Deformation potentential parameters used for $\beta$-(Al$_{x}$Ga$_{1-x}$)$_2$O$_3$ are taken from Korlacki~\textit{et al.} Ref.~\onlinecite{PhysRevApplied.18.064019}.}
\label{fig:h0lband2band}
\end{figure}
We use the obtained strain-stress relationship for the $(h0l)$ surfaces and calculate the effect of the resulting strain in the hypothetical pseudomorphically strained epitaxial $\beta$-(Al$_{x}$Ga$_{1-x}$)$_2$O$_3$ layer onto the three lowest band-to-band transitions. 
We employ the same approach proposed by Korlacki~\textit{et al.} Ref.~\onlinecite{PhysRevApplied.18.064019} which employs Vegard's rule. Using deformation potentials for $\beta$-(Al$_x$Ga$_{1-x}$)$_2$O$_3$ as described by Korlacki~\textit{et al.} Ref.~\onlinecite{PhysRevApplied.18.064019} the set of equations emerges for the combined strain and composition dependencies for transitions between $\Gamma$-point conduction band level $c-1$ and valence band level $v-1$, $v-2$, and $v-4$ ($E_{\Gamma_{1-1}}$, $E_{\Gamma_{1-2}}$, and $E_{\Gamma_{1-4}}$):
\begin{equation}
\label{eq:eg11}
\begin{split}
    E_{\Gamma_{1-1}} (\mathrm{eV}) = 5.04+2.613x\\ + (-8.38 - 0.90x) \epsilon_{\mathrm{xx}} + (0.346 + 0.22x)\epsilon_{\mathrm{xy}}\\
    +(-5.93 - 1.23x)\epsilon_{\mathrm{yy}} +(-9.56 -1.71x)\epsilon_{\mathrm{zz}}\\
    +b_{\Gamma_{1-1}}x(x-1),
\end{split}    
\end{equation}
\begin{equation}
\label{eq:eg12}
\begin{split}
    E_{\Gamma_{1-2}} (\mathrm{eV}) = 5.40+2.594x\\ + (-7.75 - 0.63x) \epsilon_{\mathrm{xx}} + (2.29 + 1.54x)\epsilon_{\mathrm{xy}}\\
    +(-8.49 -1.46x) \epsilon_{\mathrm{yy}} +(-7.74 - 1.01x)\epsilon_{\mathrm{zz}}\\
    +b_{\Gamma_{1-2}}x(x-1),
\end{split}
\end{equation}
\begin{equation}
\label{eq:eg14}
\begin{split}
    E_{\Gamma_{1-4}} (\mathrm{eV}) = 5.64+2.648x\\ + (-8.43 + 0.76x) \epsilon_{\mathrm{xx}} + (3.34 - 0.52x) \epsilon_{\mathrm{xy}}\\
    +(-11.9 - 3.41x) \epsilon_{\mathrm{yy}} + (-6.33 - 0.98x)\epsilon_{\mathrm{zz}}\\
    +b_{\Gamma_{1-4}}x(x-1),
\end{split}
\end{equation}
\noindent where $x$ is the aluminum concentration, and we ignore the bowing parameters, $b_{\Gamma_{1-1}}=0$, $b_{\Gamma_{1-2}}=0$, and $b_{\Gamma_{1-4}}=0$ for convenience. The latter are non-zero in reality but small and depend on ordering and higher-order lattice perturbations. Their effect maybe ignored here for the time being. The results of these transition energies for the three different cells are shown for fully strained $\beta$-(Al$_{x}$Ga$_{1-x}$)$_2$O$_3$ in Fig.~\ref{fig:h0lband2band}. As can be seen, the band-to-band transitions vary strongly across the different crystallographic surface orientations, and the shifts are stronger with increasing cell index. The latter is expected since the overall strain is largest for the cell with the strongest lattice distortions. It is also noted that all transitions are blue-shifted regardless of orientation and for all cells.

\section{Conclusions}

The approach introduced here reveals the existence of exactly three unit cells under a given biaxial strain situation for strain in pseudomorphic grown heteroepitaxial systems with monoclinic crystal systems. It is anticipated that in cases when the template surface has no lattice parameter within its plane, the biaxial strain will reduce the symmetry of the monoclinic epitaxial layer to triclinic and introduce shear strains and shear stresses in all elements of $\epsilon$ and $\sigma$. Then, all three unit cells may compete in epitaxial growth with stress free directions neither parallel nor perpendicular to the template surface. In this work here, we demonstrated the approach for monoclinic symmetry and a special class of surfaces, $(h0l)$. In this example, the symmetry of the epitaxial layer remains monoclinic. We find that for almost all surfaces in this class of templates two domains can exist whose stress free direction is not perpendicular to the surface. However, one domain is favored energetically and by having its stress free direction closer to the surface normal than the other. The approach presented here is generally applicable to higher symmetries, and by further modification also to triclinic materials.

\begin{acknowledgments}
This work was supported in part by the National Science Foundation (NSF) under awards NSF ECCS 2329940 and NSF/EPSCoR RII Track-1: Emergent Quantum Materials and Technologies (EQUATE), Award OIA-2044049, by Air Force Office of Scientific Research under awards FA9550-18-1-0360, FA9550-19-S-0003, FA9550-21-1-0259, and FA9550-23-1-0574 DEF, by the Knut and Alice Wallenbergs Foundation award 'Wide-bandgap semiconductors for next generation quantum components', the Swedish Energy Agency under Award No. P45396-1, the Swedish Governmental Agency for Innovation Systems (VINNOVA) under the Competence Center Program Grant No. 2016-05190, the Swedish Research Council VR under Grands No. 2016-00889, Swedish Foundation for Strategic Research under Grants No. RIF14-055, and No. EM16-0024, and the Swedish Government Strategic Research Area in Materials Science on Functional Materials at Link\"oping University, Faculty Grant SFO Mat LiU No. 2009-00971. M.S. acknowledges the University of Nebraska Foundation and the J.~A.~Woollam Foundation for support.
\end{acknowledgments}


\begin{thebibliography}{24}%
\makeatletter
\providecommand \@ifxundefined [1]{%
 \@ifx{#1\undefined}
}%
\providecommand \@ifnum [1]{%
 \ifnum #1\expandafter \@firstoftwo
 \else \expandafter \@secondoftwo
 \fi
}%
\providecommand \@ifx [1]{%
 \ifx #1\expandafter \@firstoftwo
 \else \expandafter \@secondoftwo
 \fi
}%
\providecommand \natexlab [1]{#1}%
\providecommand \enquote  [1]{``#1''}%
\providecommand \bibnamefont  [1]{#1}%
\providecommand \bibfnamefont [1]{#1}%
\providecommand \citenamefont [1]{#1}%
\providecommand \href@noop [0]{\@secondoftwo}%
\providecommand \href [0]{\begingroup \@sanitize@url \@href}%
\providecommand \@href[1]{\@@startlink{#1}\@@href}%
\providecommand \@@href[1]{\endgroup#1\@@endlink}%
\providecommand \@sanitize@url [0]{\catcode `\\12\catcode `\$12\catcode
  `\&12\catcode `\#12\catcode `\^12\catcode `\_12\catcode `\%12\relax}%
\providecommand \@@startlink[1]{}%
\providecommand \@@endlink[0]{}%
\providecommand \url  [0]{\begingroup\@sanitize@url \@url }%
\providecommand \@url [1]{\endgroup\@href {#1}{\urlprefix }}%
\providecommand \urlprefix  [0]{URL }%
\providecommand \Eprint [0]{\href }%
\providecommand \doibase [0]{http://dx.doi.org/}%
\providecommand \selectlanguage [0]{\@gobble}%
\providecommand \bibinfo  [0]{\@secondoftwo}%
\providecommand \bibfield  [0]{\@secondoftwo}%
\providecommand \translation [1]{[#1]}%
\providecommand \BibitemOpen [0]{}%
\providecommand \bibitemStop [0]{}%
\providecommand \bibitemNoStop [0]{.\EOS\space}%
\providecommand \EOS [0]{\spacefactor3000\relax}%
\providecommand \BibitemShut  [1]{\csname bibitem#1\endcsname}%
\let\auto@bib@innerbib\@empty
\bibitem [{\citenamefont {Mazzolini}\ \emph {et~al.}(2024)\citenamefont
  {Mazzolini}, \citenamefont {Wouters}, \citenamefont {Albrecht}, \citenamefont
  {Falkenstein}, \citenamefont {Martin}, \citenamefont {Vogt},\ and\
  \citenamefont {Bierwagen}}]{doi:10.1021/acsami.3c19095}%
  \BibitemOpen
  \bibfield  {author} {\bibinfo {author} {\bibfnamefont {P.}~\bibnamefont
  {Mazzolini}}, \bibinfo {author} {\bibfnamefont {C.}~\bibnamefont {Wouters}},
  \bibinfo {author} {\bibfnamefont {M.}~\bibnamefont {Albrecht}}, \bibinfo
  {author} {\bibfnamefont {A.}~\bibnamefont {Falkenstein}}, \bibinfo {author}
  {\bibfnamefont {M.}~\bibnamefont {Martin}}, \bibinfo {author} {\bibfnamefont
  {P.}~\bibnamefont {Vogt}}, \ and\ \bibinfo {author} {\bibfnamefont
  {O.}~\bibnamefont {Bierwagen}},\ }\href {\doibase 10.1021/acsami.3c19095}
  {\bibfield  {journal} {\bibinfo  {journal} {ACS Applied Materials \&
  Interfaces}\ }\textbf {\bibinfo {volume} {16}},\ \bibinfo {pages} {12793}
  (\bibinfo {year} {2024})},\ \bibinfo {note} {pMID: 38422376},\ \Eprint
  {http://arxiv.org/abs/https://doi.org/10.1021/acsami.3c19095}
  {https://doi.org/10.1021/acsami.3c19095} \BibitemShut {NoStop}%
\bibitem [{\citenamefont {Korlacki}\ \emph {et~al.}(2022)\citenamefont
  {Korlacki}, \citenamefont {Hilfiker}, \citenamefont {Knudtson}, \citenamefont
  {Stokey}, \citenamefont {Kilic}, \citenamefont {Mauze}, \citenamefont
  {Zhang}, \citenamefont {Speck}, \citenamefont {Darakchieva},\ and\
  \citenamefont {Schubert}}]{PhysRevApplied.18.064019}%
  \BibitemOpen
  \bibfield  {author} {\bibinfo {author} {\bibfnamefont {R.}~\bibnamefont
  {Korlacki}}, \bibinfo {author} {\bibfnamefont {M.}~\bibnamefont {Hilfiker}},
  \bibinfo {author} {\bibfnamefont {J.}~\bibnamefont {Knudtson}}, \bibinfo
  {author} {\bibfnamefont {M.}~\bibnamefont {Stokey}}, \bibinfo {author}
  {\bibfnamefont {U.}~\bibnamefont {Kilic}}, \bibinfo {author} {\bibfnamefont
  {A.}~\bibnamefont {Mauze}}, \bibinfo {author} {\bibfnamefont
  {Y.}~\bibnamefont {Zhang}}, \bibinfo {author} {\bibfnamefont
  {J.}~\bibnamefont {Speck}}, \bibinfo {author} {\bibfnamefont
  {V.}~\bibnamefont {Darakchieva}}, \ and\ \bibinfo {author} {\bibfnamefont
  {M.}~\bibnamefont {Schubert}},\ }\href {\doibase
  10.1103/PhysRevApplied.18.064019} {\bibfield  {journal} {\bibinfo  {journal}
  {Phys. Rev. Applied}\ }\textbf {\bibinfo {volume} {18}},\ \bibinfo {pages}
  {064019} (\bibinfo {year} {2022})}\BibitemShut {NoStop}%
\bibitem [{\citenamefont {Peelaers}\ \emph {et~al.}(2018)\citenamefont
  {Peelaers}, \citenamefont {Varley}, \citenamefont {Speck},\ and\
  \citenamefont {Van~de Walle}}]{10.1063/1.5036991}%
  \BibitemOpen
  \bibfield  {author} {\bibinfo {author} {\bibfnamefont {H.}~\bibnamefont
  {Peelaers}}, \bibinfo {author} {\bibfnamefont {J.~B.}\ \bibnamefont
  {Varley}}, \bibinfo {author} {\bibfnamefont {J.~S.}\ \bibnamefont {Speck}}, \
  and\ \bibinfo {author} {\bibfnamefont {C.~G.}\ \bibnamefont {Van~de Walle}},\
  }\href {\doibase 10.1063/1.5036991} {\bibfield  {journal} {\bibinfo
  {journal} {Applied Physics Letters}\ }\textbf {\bibinfo {volume} {112}},\
  \bibinfo {pages} {242101} (\bibinfo {year} {2018})},\ \Eprint
  {http://arxiv.org/abs/https://pubs.aip.org/aip/apl/article-pdf/doi/10.1063/1.5036991/14511675/242101\_1\_online.pdf}
  {https://pubs.aip.org/aip/apl/article-pdf/doi/10.1063/1.5036991/14511675/242101\_1\_online.pdf}
  \BibitemShut {NoStop}%
\bibitem [{\citenamefont {Liu}\ and\ \citenamefont
  {Tan}(2019)}]{10.1063/1.5093195}%
  \BibitemOpen
  \bibfield  {author} {\bibinfo {author} {\bibfnamefont {X.}~\bibnamefont
  {Liu}}\ and\ \bibinfo {author} {\bibfnamefont {C.-K.}\ \bibnamefont {Tan}},\
  }\href {\doibase 10.1063/1.5093195} {\bibfield  {journal} {\bibinfo
  {journal} {AIP Advances}\ }\textbf {\bibinfo {volume} {9}},\ \bibinfo {pages}
  {035318} (\bibinfo {year} {2019})},\ \Eprint
  {http://arxiv.org/abs/https://pubs.aip.org/aip/adv/article-pdf/doi/10.1063/1.5093195/12847300/035318\_1\_online.pdf}
  {https://pubs.aip.org/aip/adv/article-pdf/doi/10.1063/1.5093195/12847300/035318\_1\_online.pdf}
  \BibitemShut {NoStop}%
\bibitem [{\citenamefont {Peelaers}\ \emph {et~al.}(2015)\citenamefont
  {Peelaers}, \citenamefont {Steiauf}, \citenamefont {Varley}, \citenamefont
  {Janotti},\ and\ \citenamefont {Van~de Walle}}]{PhysRevB.92.085206}%
  \BibitemOpen
  \bibfield  {author} {\bibinfo {author} {\bibfnamefont {H.}~\bibnamefont
  {Peelaers}}, \bibinfo {author} {\bibfnamefont {D.}~\bibnamefont {Steiauf}},
  \bibinfo {author} {\bibfnamefont {J.~B.}\ \bibnamefont {Varley}}, \bibinfo
  {author} {\bibfnamefont {A.}~\bibnamefont {Janotti}}, \ and\ \bibinfo
  {author} {\bibfnamefont {C.~G.}\ \bibnamefont {Van~de Walle}},\ }\href
  {\doibase 10.1103/PhysRevB.92.085206} {\bibfield  {journal} {\bibinfo
  {journal} {Phys. Rev. B}\ }\textbf {\bibinfo {volume} {92}},\ \bibinfo
  {pages} {085206} (\bibinfo {year} {2015})}\BibitemShut {NoStop}%
\bibitem [{\citenamefont {{\AA}hman}\ \emph {et~al.}(1996)\citenamefont
  {{\AA}hman}, \citenamefont {Svensson},\ and\ \citenamefont
  {Albertsson}}]{Ahman:fg1144}%
  \BibitemOpen
  \bibfield  {author} {\bibinfo {author} {\bibfnamefont {J.}~\bibnamefont
  {{\AA}hman}}, \bibinfo {author} {\bibfnamefont {G.}~\bibnamefont {Svensson}},
  \ and\ \bibinfo {author} {\bibfnamefont {J.}~\bibnamefont {Albertsson}},\
  }\href {\doibase 10.1107/S0108270195016404} {\bibfield  {journal} {\bibinfo
  {journal} {Acta Crystallographica Section C}\ }\textbf {\bibinfo {volume}
  {52}},\ \bibinfo {pages} {1336} (\bibinfo {year} {1996})}\BibitemShut
  {NoStop}%
\bibitem [{\citenamefont {Vogt}\ and\ \citenamefont
  {Bierwagen}(2015)}]{10.1063/1.4913447}%
  \BibitemOpen
  \bibfield  {author} {\bibinfo {author} {\bibfnamefont {P.}~\bibnamefont
  {Vogt}}\ and\ \bibinfo {author} {\bibfnamefont {O.}~\bibnamefont
  {Bierwagen}},\ }\href {\doibase 10.1063/1.4913447} {\bibfield  {journal}
  {\bibinfo  {journal} {Applied Physics Letters}\ }\textbf {\bibinfo {volume}
  {106}},\ \bibinfo {pages} {081910} (\bibinfo {year} {2015})},\ \Eprint
  {http://arxiv.org/abs/https://pubs.aip.org/aip/apl/article-pdf/doi/10.1063/1.4913447/13592961/081910\_1\_online.pdf}
  {https://pubs.aip.org/aip/apl/article-pdf/doi/10.1063/1.4913447/13592961/081910\_1\_online.pdf}
  \BibitemShut {NoStop}%
\bibitem [{\citenamefont {Spencer}\ \emph {et~al.}(2022)\citenamefont
  {Spencer}, \citenamefont {Mock}, \citenamefont {Jacobs}, \citenamefont
  {Schubert}, \citenamefont {Zhang},\ and\ \citenamefont
  {Tadjer}}]{10.1063/5.0078037}%
  \BibitemOpen
  \bibfield  {author} {\bibinfo {author} {\bibfnamefont {J.~A.}\ \bibnamefont
  {Spencer}}, \bibinfo {author} {\bibfnamefont {A.~L.}\ \bibnamefont {Mock}},
  \bibinfo {author} {\bibfnamefont {A.~G.}\ \bibnamefont {Jacobs}}, \bibinfo
  {author} {\bibfnamefont {M.}~\bibnamefont {Schubert}}, \bibinfo {author}
  {\bibfnamefont {Y.}~\bibnamefont {Zhang}}, \ and\ \bibinfo {author}
  {\bibfnamefont {M.~J.}\ \bibnamefont {Tadjer}},\ }\href {\doibase
  10.1063/5.0078037} {\bibfield  {journal} {\bibinfo  {journal} {Applied
  Physics Reviews}\ }\textbf {\bibinfo {volume} {9}},\ \bibinfo {pages}
  {011315} (\bibinfo {year} {2022})},\ \Eprint
  {http://arxiv.org/abs/https://pubs.aip.org/aip/apr/article-pdf/doi/10.1063/5.0078037/19807888/011315\_1\_online.pdf}
  {https://pubs.aip.org/aip/apr/article-pdf/doi/10.1063/5.0078037/19807888/011315\_1\_online.pdf}
  \BibitemShut {NoStop}%
\bibitem [{\citenamefont {Bir}\ and\ \citenamefont
  {Pikus}(1974)}]{BirPikusBook}%
  \BibitemOpen
  \bibfield  {author} {\bibinfo {author} {\bibfnamefont {L.~G.}\ \bibnamefont
  {Bir}}\ and\ \bibinfo {author} {\bibfnamefont {G.~E.}\ \bibnamefont
  {Pikus}},\ }\href@noop {} {\emph {\bibinfo {title} {Symmetry and
  Strain-induced Effects in Semiconductors}}}\ (\bibinfo  {publisher} {John
  Wiley and Sons},\ \bibinfo {year} {1974})\BibitemShut {NoStop}%
\bibitem [{\citenamefont {Zhang}\ \emph
  {et~al.}(2023{\natexlab{a}})\citenamefont {Zhang}, \citenamefont {Wu},
  \citenamefont {Xing}, \citenamefont {Zhou}, \citenamefont {Zhou},
  \citenamefont {Li},\ and\ \citenamefont {Xu}}]{ZHANG2023414851}%
  \BibitemOpen
  \bibfield  {author} {\bibinfo {author} {\bibfnamefont {C.}~\bibnamefont
  {Zhang}}, \bibinfo {author} {\bibfnamefont {X.}~\bibnamefont {Wu}}, \bibinfo
  {author} {\bibfnamefont {Y.}~\bibnamefont {Xing}}, \bibinfo {author}
  {\bibfnamefont {L.}~\bibnamefont {Zhou}}, \bibinfo {author} {\bibfnamefont
  {H.}~\bibnamefont {Zhou}}, \bibinfo {author} {\bibfnamefont {S.}~\bibnamefont
  {Li}}, \ and\ \bibinfo {author} {\bibfnamefont {N.}~\bibnamefont {Xu}},\
  }\href {\doibase https://doi.org/10.1016/j.physb.2023.414851} {\bibfield
  {journal} {\bibinfo  {journal} {Physica B: Condensed Matter}\ }\textbf
  {\bibinfo {volume} {660}},\ \bibinfo {pages} {414851} (\bibinfo {year}
  {2023}{\natexlab{a}})}\BibitemShut {NoStop}%
\bibitem [{\citenamefont {Huang}\ \emph {et~al.}(2023)\citenamefont {Huang},
  \citenamefont {Johnson}, \citenamefont {Chae}, \citenamefont {Senckowski},
  \citenamefont {Wong},\ and\ \citenamefont {Hwang}}]{10.1063/5.0156009}%
  \BibitemOpen
  \bibfield  {author} {\bibinfo {author} {\bibfnamefont {H.-L.}\ \bibnamefont
  {Huang}}, \bibinfo {author} {\bibfnamefont {J.~M.}\ \bibnamefont {Johnson}},
  \bibinfo {author} {\bibfnamefont {C.}~\bibnamefont {Chae}}, \bibinfo {author}
  {\bibfnamefont {A.}~\bibnamefont {Senckowski}}, \bibinfo {author}
  {\bibfnamefont {M.~H.}\ \bibnamefont {Wong}}, \ and\ \bibinfo {author}
  {\bibfnamefont {J.}~\bibnamefont {Hwang}},\ }\href {\doibase
  10.1063/5.0156009} {\bibfield  {journal} {\bibinfo  {journal} {Applied
  Physics Letters}\ }\textbf {\bibinfo {volume} {122}},\ \bibinfo {pages}
  {251602} (\bibinfo {year} {2023})},\ \Eprint
  {http://arxiv.org/abs/https://pubs.aip.org/aip/apl/article-pdf/doi/10.1063/5.0156009/18006597/251602\_1\_5.0156009.pdf}
  {https://pubs.aip.org/aip/apl/article-pdf/doi/10.1063/5.0156009/18006597/251602\_1\_5.0156009.pdf}
  \BibitemShut {NoStop}%
\bibitem [{\citenamefont {Seacat}\ \emph {et~al.}(2024)\citenamefont {Seacat},
  \citenamefont {Lyons},\ and\ \citenamefont
  {Peelaers}}]{PhysRevMaterials.8.014601}%
  \BibitemOpen
  \bibfield  {author} {\bibinfo {author} {\bibfnamefont {S.}~\bibnamefont
  {Seacat}}, \bibinfo {author} {\bibfnamefont {J.~L.}\ \bibnamefont {Lyons}}, \
  and\ \bibinfo {author} {\bibfnamefont {H.}~\bibnamefont {Peelaers}},\ }\href
  {\doibase 10.1103/PhysRevMaterials.8.014601} {\bibfield  {journal} {\bibinfo
  {journal} {Phys. Rev. Mater.}\ }\textbf {\bibinfo {volume} {8}},\ \bibinfo
  {pages} {014601} (\bibinfo {year} {2024})}\BibitemShut {NoStop}%
\bibitem [{\citenamefont {Barmore}\ \emph {et~al.}(2023)\citenamefont
  {Barmore}, \citenamefont {Jesenovec}, \citenamefont {McCloy},\ and\
  \citenamefont {McCluskey}}]{10.1063/5.0149900}%
  \BibitemOpen
  \bibfield  {author} {\bibinfo {author} {\bibfnamefont {L.~M.}\ \bibnamefont
  {Barmore}}, \bibinfo {author} {\bibfnamefont {J.}~\bibnamefont {Jesenovec}},
  \bibinfo {author} {\bibfnamefont {J.~S.}\ \bibnamefont {McCloy}}, \ and\
  \bibinfo {author} {\bibfnamefont {M.~D.}\ \bibnamefont {McCluskey}},\ }\href
  {\doibase 10.1063/5.0149900} {\bibfield  {journal} {\bibinfo  {journal}
  {Journal of Applied Physics}\ }\textbf {\bibinfo {volume} {133}},\ \bibinfo
  {pages} {175703} (\bibinfo {year} {2023})},\ \Eprint
  {http://arxiv.org/abs/https://pubs.aip.org/aip/jap/article-pdf/doi/10.1063/5.0149900/17304220/175703\_1\_5.0149900.pdf}
  {https://pubs.aip.org/aip/jap/article-pdf/doi/10.1063/5.0149900/17304220/175703\_1\_5.0149900.pdf}
  \BibitemShut {NoStop}%
\bibitem [{\citenamefont {Xu}\ \emph {et~al.}(2019)\citenamefont {Xu},
  \citenamefont {Park}, \citenamefont {Yao}, \citenamefont {Wolverton},
  \citenamefont {Razeghi}, \citenamefont {Wu},\ and\ \citenamefont
  {Dravid}}]{doi:10.1021/acsami.8b17731}%
  \BibitemOpen
  \bibfield  {author} {\bibinfo {author} {\bibfnamefont {Y.}~\bibnamefont
  {Xu}}, \bibinfo {author} {\bibfnamefont {J.-H.}\ \bibnamefont {Park}},
  \bibinfo {author} {\bibfnamefont {Z.}~\bibnamefont {Yao}}, \bibinfo {author}
  {\bibfnamefont {C.}~\bibnamefont {Wolverton}}, \bibinfo {author}
  {\bibfnamefont {M.}~\bibnamefont {Razeghi}}, \bibinfo {author} {\bibfnamefont
  {J.}~\bibnamefont {Wu}}, \ and\ \bibinfo {author} {\bibfnamefont {V.~P.}\
  \bibnamefont {Dravid}},\ }\href {\doibase 10.1021/acsami.8b17731} {\bibfield
  {journal} {\bibinfo  {journal} {ACS Applied Materials \& Interfaces}\
  }\textbf {\bibinfo {volume} {11}},\ \bibinfo {pages} {5536} (\bibinfo {year}
  {2019})}\BibitemShut {NoStop}%
\bibitem [{\citenamefont {Zhang}\ \emph
  {et~al.}(2023{\natexlab{b}})\citenamefont {Zhang}, \citenamefont {Li},
  \citenamefont {Wu}, \citenamefont {Li}, \citenamefont {Zhang}, \citenamefont
  {Liang},\ and\ \citenamefont {Shen}}]{ZHANG2023106916}%
  \BibitemOpen
  \bibfield  {author} {\bibinfo {author} {\bibfnamefont {R.}~\bibnamefont
  {Zhang}}, \bibinfo {author} {\bibfnamefont {M.}~\bibnamefont {Li}}, \bibinfo
  {author} {\bibfnamefont {G.}~\bibnamefont {Wu}}, \bibinfo {author}
  {\bibfnamefont {L.}~\bibnamefont {Li}}, \bibinfo {author} {\bibfnamefont
  {Z.}~\bibnamefont {Zhang}}, \bibinfo {author} {\bibfnamefont
  {K.}~\bibnamefont {Liang}}, \ and\ \bibinfo {author} {\bibfnamefont
  {W.}~\bibnamefont {Shen}},\ }\href {\doibase
  https://doi.org/10.1016/j.rinp.2023.106916} {\bibfield  {journal} {\bibinfo
  {journal} {Results in Physics}\ }\textbf {\bibinfo {volume} {52}},\ \bibinfo
  {pages} {106916} (\bibinfo {year} {2023}{\natexlab{b}})}\BibitemShut
  {NoStop}%
\bibitem [{\citenamefont {Hara}\ \emph {et~al.}(2023)\citenamefont {Hara},
  \citenamefont {Zhu}, \citenamefont {Deng}, \citenamefont {Marin},
  \citenamefont {Guo},\ and\ \citenamefont {Pezzotti}}]{Hara_2023}%
  \BibitemOpen
  \bibfield  {author} {\bibinfo {author} {\bibfnamefont {Y.}~\bibnamefont
  {Hara}}, \bibinfo {author} {\bibfnamefont {W.}~\bibnamefont {Zhu}}, \bibinfo
  {author} {\bibfnamefont {G.}~\bibnamefont {Deng}}, \bibinfo {author}
  {\bibfnamefont {E.}~\bibnamefont {Marin}}, \bibinfo {author} {\bibfnamefont
  {Q.}~\bibnamefont {Guo}}, \ and\ \bibinfo {author} {\bibfnamefont
  {G.}~\bibnamefont {Pezzotti}},\ }\href {\doibase 10.1088/1361-6463/acbbdb}
  {\bibfield  {journal} {\bibinfo  {journal} {Journal of Physics D: Applied
  Physics}\ }\textbf {\bibinfo {volume} {56}},\ \bibinfo {pages} {125102}
  (\bibinfo {year} {2023})}\BibitemShut {NoStop}%
\bibitem [{\citenamefont {Uchida}\ and\ \citenamefont
  {Sugie}(2023)}]{Uchida_2023}%
  \BibitemOpen
  \bibfield  {author} {\bibinfo {author} {\bibfnamefont {T.}~\bibnamefont
  {Uchida}}\ and\ \bibinfo {author} {\bibfnamefont {R.}~\bibnamefont {Sugie}},\
  }\href {\doibase 10.35848/1347-4065/acb26f} {\bibfield  {journal} {\bibinfo
  {journal} {Japanese Journal of Applied Physics}\ }\textbf {\bibinfo {volume}
  {62}},\ \bibinfo {pages} {SF1003} (\bibinfo {year} {2023})}\BibitemShut
  {NoStop}%
\bibitem [{\citenamefont {Korlacki}\ \emph {et~al.}(2020)\citenamefont
  {Korlacki}, \citenamefont {Stokey}, \citenamefont {Mock}, \citenamefont
  {Knight}, \citenamefont {Papamichail}, \citenamefont {Darakchieva},\ and\
  \citenamefont {Schubert}}]{PhysRevB.102.180101}%
  \BibitemOpen
  \bibfield  {author} {\bibinfo {author} {\bibfnamefont {R.}~\bibnamefont
  {Korlacki}}, \bibinfo {author} {\bibfnamefont {M.}~\bibnamefont {Stokey}},
  \bibinfo {author} {\bibfnamefont {A.}~\bibnamefont {Mock}}, \bibinfo {author}
  {\bibfnamefont {S.}~\bibnamefont {Knight}}, \bibinfo {author} {\bibfnamefont
  {A.}~\bibnamefont {Papamichail}}, \bibinfo {author} {\bibfnamefont
  {V.}~\bibnamefont {Darakchieva}}, \ and\ \bibinfo {author} {\bibfnamefont
  {M.}~\bibnamefont {Schubert}},\ }\href {\doibase 10.1103/PhysRevB.102.180101}
  {\bibfield  {journal} {\bibinfo  {journal} {Phys. Rev. B}\ }\textbf {\bibinfo
  {volume} {102}},\ \bibinfo {pages} {180101} (\bibinfo {year}
  {2020})}\BibitemShut {NoStop}%
\bibitem [{\citenamefont {Hasuike}\ \emph {et~al.}(2023)\citenamefont
  {Hasuike}, \citenamefont {Maeda}, \citenamefont {Isaji}, \citenamefont
  {Kobayashi}, \citenamefont {Ohira},\ and\ \citenamefont
  {Isshiki}}]{Hasuike_2023}%
  \BibitemOpen
  \bibfield  {author} {\bibinfo {author} {\bibfnamefont {N.}~\bibnamefont
  {Hasuike}}, \bibinfo {author} {\bibfnamefont {I.}~\bibnamefont {Maeda}},
  \bibinfo {author} {\bibfnamefont {S.}~\bibnamefont {Isaji}}, \bibinfo
  {author} {\bibfnamefont {K.}~\bibnamefont {Kobayashi}}, \bibinfo {author}
  {\bibfnamefont {K.}~\bibnamefont {Ohira}}, \ and\ \bibinfo {author}
  {\bibfnamefont {T.}~\bibnamefont {Isshiki}},\ }\href {\doibase
  10.35848/1347-4065/acc74a} {\bibfield  {journal} {\bibinfo  {journal}
  {Japanese Journal of Applied Physics}\ }\textbf {\bibinfo {volume} {62}},\
  \bibinfo {pages} {SF1020} (\bibinfo {year} {2023})}\BibitemShut {NoStop}%
\bibitem [{\citenamefont {Mu}\ \emph {et~al.}(2020)\citenamefont {Mu},
  \citenamefont {Peelaers}, \citenamefont {Zhang}, \citenamefont {Wang},\ and\
  \citenamefont {Van~de Walle}}]{10.1063/5.0036072}%
  \BibitemOpen
  \bibfield  {author} {\bibinfo {author} {\bibfnamefont {S.}~\bibnamefont
  {Mu}}, \bibinfo {author} {\bibfnamefont {H.}~\bibnamefont {Peelaers}},
  \bibinfo {author} {\bibfnamefont {Y.}~\bibnamefont {Zhang}}, \bibinfo
  {author} {\bibfnamefont {M.}~\bibnamefont {Wang}}, \ and\ \bibinfo {author}
  {\bibfnamefont {C.~G.}\ \bibnamefont {Van~de Walle}},\ }\href {\doibase
  10.1063/5.0036072} {\bibfield  {journal} {\bibinfo  {journal} {Applied
  Physics Letters}\ }\textbf {\bibinfo {volume} {117}},\ \bibinfo {pages}
  {252104} (\bibinfo {year} {2020})},\ \Eprint
  {http://arxiv.org/abs/https://pubs.aip.org/aip/apl/article-pdf/doi/10.1063/5.0036072/14541856/252104\_1\_online.pdf}
  {https://pubs.aip.org/aip/apl/article-pdf/doi/10.1063/5.0036072/14541856/252104\_1\_online.pdf}
  \BibitemShut {NoStop}%
\bibitem [{\citenamefont
  {Grundmann}(2017)}]{https://doi.org/10.1002/pssb.201700134}%
  \BibitemOpen
  \bibfield  {author} {\bibinfo {author} {\bibfnamefont {M.}~\bibnamefont
  {Grundmann}},\ }\href {\doibase https://doi.org/10.1002/pssb.201700134}
  {\bibfield  {journal} {\bibinfo  {journal} {physica status solidi (b)}\
  }\textbf {\bibinfo {volume} {254}} (\bibinfo {year} {2017}),\
  https://doi.org/10.1002/pssb.201700134},\ \Eprint
  {http://arxiv.org/abs/https://onlinelibrary.wiley.com/doi/pdf/10.1002/pssb.201700134}
  {https://onlinelibrary.wiley.com/doi/pdf/10.1002/pssb.201700134} \BibitemShut
  {NoStop}%
\bibitem [{\citenamefont {Schubert}\ \emph {et~al.}(2016)\citenamefont
  {Schubert}, \citenamefont {Korlacki}, \citenamefont {Knight}, \citenamefont
  {Hofmann}, \citenamefont {Sch\"oche}, \citenamefont {Darakchieva},
  \citenamefont {Janz\'en}, \citenamefont {Monemar}, \citenamefont {Gogova},
  \citenamefont {Thieu}, \citenamefont {Togashi}, \citenamefont {Murakami},
  \citenamefont {Kumagai}, \citenamefont {Goto}, \citenamefont {Kuramata},
  \citenamefont {Yamakoshi},\ and\ \citenamefont
  {Higashiwaki}}]{PhysRevB.93.125209}%
  \BibitemOpen
  \bibfield  {author} {\bibinfo {author} {\bibfnamefont {M.}~\bibnamefont
  {Schubert}}, \bibinfo {author} {\bibfnamefont {R.}~\bibnamefont {Korlacki}},
  \bibinfo {author} {\bibfnamefont {S.}~\bibnamefont {Knight}}, \bibinfo
  {author} {\bibfnamefont {T.}~\bibnamefont {Hofmann}}, \bibinfo {author}
  {\bibfnamefont {S.}~\bibnamefont {Sch\"oche}}, \bibinfo {author}
  {\bibfnamefont {V.}~\bibnamefont {Darakchieva}}, \bibinfo {author}
  {\bibfnamefont {E.}~\bibnamefont {Janz\'en}}, \bibinfo {author}
  {\bibfnamefont {B.}~\bibnamefont {Monemar}}, \bibinfo {author} {\bibfnamefont
  {D.}~\bibnamefont {Gogova}}, \bibinfo {author} {\bibfnamefont {Q.-T.}\
  \bibnamefont {Thieu}}, \bibinfo {author} {\bibfnamefont {R.}~\bibnamefont
  {Togashi}}, \bibinfo {author} {\bibfnamefont {H.}~\bibnamefont {Murakami}},
  \bibinfo {author} {\bibfnamefont {Y.}~\bibnamefont {Kumagai}}, \bibinfo
  {author} {\bibfnamefont {K.}~\bibnamefont {Goto}}, \bibinfo {author}
  {\bibfnamefont {A.}~\bibnamefont {Kuramata}}, \bibinfo {author}
  {\bibfnamefont {S.}~\bibnamefont {Yamakoshi}}, \ and\ \bibinfo {author}
  {\bibfnamefont {M.}~\bibnamefont {Higashiwaki}},\ }\href {\doibase
  10.1103/PhysRevB.93.125209} {\bibfield  {journal} {\bibinfo  {journal} {Phys.
  Rev. B}\ }\textbf {\bibinfo {volume} {93}},\ \bibinfo {pages} {125209}
  (\bibinfo {year} {2016})}\BibitemShut {NoStop}%
\bibitem [{\citenamefont {Kranert}\ \emph {et~al.}(2015)\citenamefont
  {Kranert}, \citenamefont {Jenderka}, \citenamefont {Lenzner}, \citenamefont
  {Lorenz}, \citenamefont {von Wenckstern}, \citenamefont {Schmidt-Grund},\
  and\ \citenamefont {Grundmann}}]{doi:10.1063/1.4915627}%
  \BibitemOpen
  \bibfield  {author} {\bibinfo {author} {\bibfnamefont {C.}~\bibnamefont
  {Kranert}}, \bibinfo {author} {\bibfnamefont {M.}~\bibnamefont {Jenderka}},
  \bibinfo {author} {\bibfnamefont {J.}~\bibnamefont {Lenzner}}, \bibinfo
  {author} {\bibfnamefont {M.}~\bibnamefont {Lorenz}}, \bibinfo {author}
  {\bibfnamefont {H.}~\bibnamefont {von Wenckstern}}, \bibinfo {author}
  {\bibfnamefont {R.}~\bibnamefont {Schmidt-Grund}}, \ and\ \bibinfo {author}
  {\bibfnamefont {M.}~\bibnamefont {Grundmann}},\ }\href {\doibase
  10.1063/1.4915627} {\bibfield  {journal} {\bibinfo  {journal} {Journal of
  Applied Physics}\ }\textbf {\bibinfo {volume} {117}},\ \bibinfo {pages}
  {125703} (\bibinfo {year} {2015})},\ \Eprint
  {http://arxiv.org/abs/https://doi.org/10.1063/1.4915627}
  {https://doi.org/10.1063/1.4915627} \BibitemShut {NoStop}%
\bibitem [{Note1()}]{Note1}%
  \BibitemOpen
  \bibinfo {note} {Note the structure of the determinant for the stress tensor
  where component $\sigma _{zz}$ always factorizes out. This has trivial
  solutions then for all cases when the $b$ axis is perpendicular to the
  surface, (010). Then the $z$ direction is also parallel to the surface normal
  $\protect \mathbf {g}_{hkl}$. Hence, for growth on $(010)$ only one stress
  free solution with a stress free direction not parallel to the surface exists
  which is trivially then the surface normal. For almost all other
  crystallographic surfaces, the stress free directions are neither parallel
  nor perpendicular to the surface and all represent potential stable unit
  cells during growth.}\BibitemShut {Stop}%
\end{thebibliography}
%
\end{document}